\documentclass[twocolumn]{aastex631}

\usepackage{amsmath}
\shorttitle{The Equilibrium Tide}
\shortauthors{Preece et al.}
\graphicspath{{./}{figures/}}

\begin{document}

\title{The Equilibrium Tide: An Updated Prescription for Population Synthesis Codes}

\author[0000-0001-7984-7033]{Holly P. Preece}
\affiliation{Max Plank Institute for  Astrophysics}

\author[0000-0003-1004-5635]{Adrian S. Hamers}
\affiliation{Max Plank Institute for  Astrophysics}

\author[0000-0001-5853-6017]{Patrick G. Neunteufel}
\affiliation{Max Plank Institute for  Astrophysics}

\author{Adam L. Schaefer}
\affiliation{Max Plank Institute for  Astrophysics}

\author[0000-0002-1556-9449]{Christopher A. Tout}
\affiliation{Institute of Astronomy, University of Cambridge}



\begin{abstract}

We present an updated prescription for the equilibrium tides suitable for population synthesis codes. A grid of 1D evolutionary models was created and the viscous time-scale was calculated for each detailed model. A metallicity dependent power-law relation was fitted to both the convective cores and convective envelopes of the models. The prescription was implemented into the population synthesis code BSE and predicts an 16.5\% reduction in the overall  number of merges, with those involving main-sequence stars most affected. The new prescription also reduces the overall supernova rate by 3.6\% with individual channels being differently affected. The single degenerate Ia supernova occurrence is reduced by 12.8\%. The merging of two Carbon Oxygen white dwarfs to cause a Ia supernova occurs 16\% less frequently. The number of sub-synchronously rotating stars in close binaries is substantially increased with our prescription, as is the number of non-circularized systems at the start of common-envelope evolution.

\end{abstract}



\section{Introduction} \label{sec:intro}
Observational evidence suggests that a large fraction of stars are formed as components of binary or higher order multiples. Higher mass stars are far more likely to be in binaries. About 80\% of O stars are in binary or higher order systems and have 1.3 companions on average. Meanwhile, only 20\% of M type stars are in multiples \citep{multiplicity1,multiplicity2}. These stars are gravitationally bound to one another in elliptical orbits. If the orbits are dynamically stable they are Keplerian on short time-scales. 

 Roche surfaces are gravitational equipotential surfaces in binary systems in the frame co-rotating with the star \citep{kopalroche}. If the radius of one of the stars is greater than the Roche radius \citep{paczyroche, eggroche} mass is transferred to the companion. Binaries are described as non-interacting if they are sufficiently wide that neither of the objects' Roche lobes overflow at any evolutionary stage. These wide binaries evolve according to single evolution. Binary-star interactions between closely orbiting bodies can have a significant effect upon the evolution of stars. In addition to stable mass transfer, unstable mass transfer, common-envelope evolution and stellar merges all occur amongst close binaries. In some cases, referred to as the Algol paradox \citep{hoyle55,crawfordalgol,algol}, sufficient mass can be transferred from the more massive primary to the less massive secondary such that the secondary becomes the more massive star. Unstable mass transfer \citep{masstrasfer}can trigger common envelope evolution which shrinks the orbit and ejects the outer layers of the object. Short period binaries can produce gravitational wave sources such as black hole-neutron star merges \citep{selmabh,gwsource}. Close binaries with white dwarfs can produce Ia supernovae \citep{webbinksne,sne1a,sne1aii} and novae.
 
 Tidal interactions are slow, non-conservative processes which affect the rotation, eccentricity and inclination of close binary or higher order multiple star systems \citep{darwin,alexander,hut}. Tides act to synchronize, circularize and align rotational and orbital axes of the interacting objects. Tidal interactions cause the stars to become deformed from their spherically symmetric shape into prolate ellipsoids. Tides can shrink orbits and thus trigger mass-transfer. 

 In hierarchical triple star systems with highly inclined outer tertiary components von-Zeipel Lidov Kozai (ZLK) oscillations \citep{vonzeipel,lidov,kozai} periodically excite the eccentricity of the inner binary. The combination of tidal dissipation and ZLK oscillations is a mechanism to produce close period inner binaries \citep{zkltides,eggzkltides,eggzkltides2,fabzkltides}.

\subsection{Tidal Interactions}

\cite{darwin} formulated the earliest robust theory of tidal interactions. This theory suggested that tidal locking was achieved purely by the torque created by the tidal bulge. Unfortunately this mechanism failed to produce the torque necessary to tidally lock a stellar system in all observed cases. In convective regions the bulk movement of material over large distances causes a natural turbulent viscosity. Viscosity provides a drag which prevents the bulge moving instantaneously around the star and offers a mechanism to dissipate energy via the equilibrium tide. Among others, \citep{eggzbook,eggtides} developed a formalism to describe convective dissipation. \cite{eggtides}'s theory is self consistent and is derived from first principles using only the Navier-Stokes equation, the Poisson equation and the equation of continuity but requires a local viscosity to dissipate energy. 

There are currently two dominant classical theories of tidal interactions which attempt to answer the question of how the tidal energy is dissipated in the radiative regions. One was proposed by \cite{zahn75,zahn77} and the other by \cite{tassoul80s}. Zahn's theory of dynamical tides applies to stars with convective cores and radiative envelopes and suggests that the periodically varying potential in the star resonates with and excites the star's natural modes of oscillation. These oscillations are excited near the convective core boundary then damped in the radiative envelope which provides a dissipative mechanism for the tides. \cite{zahn77} attempts to average over the resonant modes to find the overall effect of the dynamical tide. This theory predicts reasonable circularization time-scales but synchronization time-scales which are too long to account for the observed numbers of locked systems. \cite{tassoul80s}'s hydrodynamical mechanism was proposed in an attempt to counteract these problems. It successfully predicts shorter synchronization time-scales. The tidal disruption gives rise to larger scale meridional flows. Mass exchange between an Ekman boundary layer and the rest of the star allows angular momentum exchange which can spin up or spin down the star. In an Ekman layer the pressure, the Coriolis force and the turbulent drag are balanced. \cite{tassoul80s} suggests that large-scale meridional flow very efficiently synchronizes a star. However, \cite{tassoulbad} highly contest \cite{tassoul80s}'s theory by showing that incorrect boundary conditions were used to increase the efficiency of the Ekman pumping. 

In recent years there has been substantial development of tidal theory. \cite{fuller} have successfully applied the dynamical tide to describe the long term orbital evolution of degenerate white dwarfs in binaries. In contrast to \cite{zahn77}, they calculated the individual mode resonances as opposed to the average effect. \cite{vicklai} calculate the forcing frequencies in convective regions of highly eccentric binaries. They find that at high eccentricities the tidal effect is oscillatory. At low eccentricities the weak-friction approximation of the classical equilibrium tide is recovered. \cite{terquem} and \cite{termquemii} presents a novel theory of the equilibrium tide in the fast tide regime wherein the tidal flow is treated as a rapidly fluctuating flow and the convective flow of the material in the star is treated as the mean flow. The validity of the \cite{terquem} mechanism is debated by \cite{barker}.

Precise implementation of tidal dissipation in population synthesis codes can be used to test the validity of classical tidal theory and probe any regions in the Hertzsprung-Russel diagram in which the theory breaks down. It is known that mixing length is not truly reflective of the physical processes occurring in the star. \cite{alexander}, \cite{eggtides} and \cite{hut} suggest that either the convective viscosity or the dissipation strength can be observable quantities. The viscosity is subject to the largest uncertainties.

\subsection{Population Synthesis Codes}
Population synthesis codes are used to rapidly calculate the properties of stars and stellar systems. They do not resolve the full structure of the star but instead rely on analytical fits or interpolations of evolution tracks from detailed stellar evolution codes. Because population synthesis codes are fast, they can be used to generate sufficiently large data sets to statistically represent distributions of populations of stars.

\subsubsection{BSE}

The population synthesis code Binary Star Evolution (BSE, \cite{bse}) has been used as the foundation for many subsequent binary population synthesis codes since its inception. BSE uses the  Single Star Evolution (SSE, \cite{sse}) analytic fits for stellar evolution then includes prescriptions for wind accretion, orbital changes owing to mass variations, tidal evolution, gravitational radiation and magnetic breaking, supernovae kicks, Roche lobe overflow, common envelope evolution and merging. Multiple Stellar Evolution (MSE, \cite{mse}) builds on the SSE and BSE routines to model the single, binary and dynamical evolution of multiple star systems.

In convective regions BSE uses \cite{hut}'s formalism for the equilibrium tide to calculate the time evolution of the eccentricity, semi-major axis and rotational angular velocity. The tidal coupling is approximated by \cite{bse}.

\subsection{Paper Outline}
Section 2 outlines the tidal theory used in this work, section 3 describes the stellar models used, section 4 approximates the structure dependent tidal quantities with parameters available in BSE, section 5 implements the new prescription in BSE to assess the effects of the updated prescription, section 6 is a discussion of the implications and section 7 concludes.

\section{Theory}
A derivation for the equilibrium tide from first principles was presented by \cite{eggtides, eggzbook}. The derivation assumes only that the rate of dissipation of energy should be a positive definite function of the rate of change of the tide, as viewed in a frame which rotates with the star, and that the total angular momentum is
conserved. One of the clear advantages of the formalism is that for a given 1D stellar model the tidal dissipation rate can be self-consistently calculated. The theory assumes that the lag time of the bulge is related to the quadrupole moment of the star. 
First, the magnitude of the radial distortion $\alpha$ can be found by solving

\begin{equation}
    \alpha'' - \frac{6\alpha}{r^2} + \frac{2rm'}{m}\bigg(\frac{\alpha '}{r} + \frac{\alpha}{r^2}\bigg)  = 0
\end{equation}
where $r$ is the radius and $m$ is the mass. Primes denote derivatives with respect to $r$. To first order $\alpha$ depends only on the zeroth order, spherical, structure of a non-rotating star and the value of $\alpha$ at the surface. The structure constant, $Q$ is 
 \begin{equation}
    Q = \frac{1}{5MR^2 \alpha(R)}\int_0^Mr^2(5\alpha + r\alpha ')dm
\end{equation}
where $M$ and $R$ are the total mass and radius of the star respectively. The viscous time-scale, $t_{\rm{visc}}$ is a dissipative time-scale intrinsic to the star and is defined as
\begin{equation}
    \frac{1}{t_{\rm{visc}}} = \frac{1}{M R^2} \int_{m_{\rm{c, in}}} ^{m_{\rm{c, out}}}\nu \gamma(r) dm  
\end{equation}
where $\nu$ is the turbulent viscosity of the convective region, which can be approximated by $\nu=wl/3$ \citep{zahnvisc} where $w$ is the mixing velocity and $l$ is the mixing length as predicted by \cite{mlt}. The limits of integration $m_{\rm{c, in}}$ and $m_{\rm{c, out}}$ are the inner and outer mass co-ordinates of the convective region. The $M$ and $R$ are the mass and radius of the star respectively. The factor $\gamma(r)$ is 
\begin{equation}
    \gamma(r) = \beta^2 + \frac{2}{3}r\beta \beta' + \frac{7}{30}r^2\beta'^2
\end{equation}
and relates to the integral of the square of the rate of strain tensor and $\beta$ satisfies the differential equation

\begin{equation}
 \frac{d (\rho \beta)}{dr} = \frac{\alpha(r)}{\alpha(R)}\frac{d \rho}{d \beta}
\end{equation}
At the surface $\alpha(R)$ can be determined from the strength of the perturbation. However, when calculating the tidal dissipation strength all equations have factors of $\alpha(r)/\alpha(R)$ and thus it is not necessary for this work to calculate the magnitude of the distortion.

\cite{hut}'s formalism gives the same basic result as \cite{eggtides}'s. However \cite{hut} assumes the tidal lag time is constant as in \cite{darwin} and \cite{alexander}. The resulting equations of motion are 
\begin{equation}
\begin{split}
    \frac{da}{dt} = -6 \bigg(\frac{k}{T}\bigg)_{\rm{c}} q(1+q)\bigg(\frac{R}{a}\bigg)^8\frac{a}{(1-e^2)^{15/2}} \\ \times \bigg[ f_1(e^2)-(1-e^2)^{3/2}f_2(e^2)\frac{\Omega}{\omega}\bigg],
\end{split}
\end{equation}
where $a$ is the semi-major axis, $k$ is the apsidal motion constant, $T$ is the tidal response time, $q$ is the mass ratio ($m_2/m_1)$, $e$ is the eccentricity, $\Omega$ is the spin angular frequency and $\omega$ is the orbital angular frequency,
\begin{equation}
\begin{split}
        \frac{de}{dt} = -27 \bigg(\frac{k}{T}\bigg)_{\rm{c}}q(1+q)\bigg(\frac{R}{a}\bigg)^8\frac{e}{(1-e^2)^{13/2}} \\ \times \bigg[f_3(e^2)-\frac{11}{18}(1-e^2)^{3/2}f_4(e^2)\frac{\Omega}{\omega}\bigg]
\end{split}
\end{equation}
and
\begin{equation}
    \begin{split}
        \frac{d\Omega}{dt} = 3\bigg(\frac{k}{T}\bigg)_{\rm{c}}\frac{q^2}{r_g^2}\bigg(\frac{R}{a}\bigg)^6 \frac{n}{(1-e^2)^6} \\ \times \bigg[f_2(e^2)-(1-e^2)^{3/2}f_5(e^2)\frac{\Omega}{\omega}\bigg],
    \end{split}
\end{equation}
where $r_g$ is the radius of gyration and
\begin{equation}
    f_1(e^2)=1+\frac{31}{2}e^2+\frac{255}{8}e^4+\frac{185}{16}e^6+\frac{25}{64}e^8,
\end{equation}
\begin{equation}
    f_2(e^2)=1+\frac{15}{2}e^2+\frac{45}{8}e^4+\frac{5}{16}e^6,
\end{equation}
\begin{equation}
    f_3(e^2) = 1+\frac{15}{4}e^2 + \frac{15}{8}e^4 + \frac{5}{64}e^6,
\end{equation}
\begin{equation}
    f_4(e^2)= 1+\frac{3}{2}e^2 + \frac{1}{8}e^4
\end{equation}
and
\begin{equation}
    f_5(e^2)=1+3e^2 + \frac{3}{8}e^4.
\end{equation}
 The resulting equations of motion of \cite{hut} and \cite{eggtides} have very similar forms so equating $\dot{\Omega}/\Omega$ at $e=0$ and cancelling the mutual terms gives 

\begin{equation}
    \bigg(\frac{k}{T}\bigg)_{\rm{c}} = \frac{3}{t_{\rm{visc}}} \frac{1}{(1-Q)^2}.
    \label{eq:equivalence}
\end{equation}

\subsection{BSE Implementation of the Equilibrium Tide}
The dependence on the structure of the star is contained within the $(k/T)_{\rm{c}}$ term and all other parameters are available in BSE. Following \cite{rasio}, $(k/T)_{\rm{c}}$ is approximated as 
    \begin{equation}
        \bigg( \frac{k}{T} \bigg)_{\rm{c}} = \frac{2}{21}\frac{f_{\rm{conv}}}{t_{\rm{conv}}}\frac{M_{\rm{conv}}}{M}
    \end{equation}
where $k$ is the apsidal motion constant (not to be confused with the Love number) and $T$ is the tidal response time. The convective turnover time is 
    
    \begin{equation}
        t_{\rm{conv}} = 0.4311 \bigg[ \frac{M_{\rm{conv}} R_{\rm{conv}} (R - 0.5 R_{\rm{conv}})}{3L} \bigg]^{1/3}\,\rm{yr},
        \label{eqtconv}
    \end{equation}
where $M_{\rm{conv}}$ is the mass contained within the convective envelope, $R_{\rm{conv}}$ is its radial thickness, $R$ is the total radius, $M$ is the total mass and $L$ is the luminosity. Solar units are used for all quantities. $R_{\rm{conv}}$ is defined as $r_a -r_b$ where $r_a$ and $r_b$ are the radial co-ordinates of the surface of the convective region considered and the base. The factor correcting for fast tides $f_{\rm{conv}}$ is
    
    \begin{equation}
       f_{\rm{conv}} = \rm{min} \bigg[1,\bigg(\frac{\it{P}_{\rm{tid}}}{2\it{t}_{\rm{conv}}}\bigg)^2 \bigg]
    \end{equation}
where the tidal pumping scale $P_{\rm{tid}}$ is
    
    \begin{equation}
        \frac{1}{P_{\rm{tid}}} = \bigg|\frac{1}{\it{P}_{\rm{orb}}} - \frac{1}{P_{\rm{spin}}} \bigg|.
    \end{equation}

\subsection{Fast Tides}
Fast tides are defined as structural, and hence often evolutionary, phases where the tidal period is shorter than the convective turnover time. In the fast tide regime the efficiency of tidal dissipation via the equilibrium tide is reduced. The appropriate factor for reduction of efficiency of the tides has been an area of much debate. \cite{zahn66} introduced 
\begin{equation}
       f_{\rm{conv}} = \rm{min} \bigg[1,\bigg(\frac{\it{P}_{\rm{tid}}}{2\it{t}_{\rm{conv}}}\bigg)^1 \bigg]
    \end{equation}
which corrects for the distance that the convective material moves in half an orbital period. Following \cite{chrissuggestion} \cite{bse} define the corrective factor as

\begin{equation}
       f_{\rm{conv}} = \rm{min} \bigg[1,\bigg(\frac{\it{P}_{\rm{tid}}}{2\it{t}_{\rm{conv}}}\bigg)^2 \bigg].
\end{equation}
Recent results from 3D hydro-dynamical simulations suggest a broken power law is most accurate \citep{fasttides}. The effect of different corrective factors is not considered in this work.

\section{Stellar Models}
The stellar models used for this project were created with the STARS code initially developed by \cite{eggz71, starssemiconv} and subsequently updated by \cite{chrisupdate}, \cite{starsdov} and \cite{stancliffe09}. \cite{starssemiconv} implemented semiconvection in STARS as a diffusive process which follows \cite{schwarzscsemiconv}'s prescription. It assumes that the energy transport by convection in the semiconvective region is borderline negligible but that there is substantial chemical mixing which avoids any discontinuity in the chemical profile. Semiconvective regions then have $\nabla_{\rm{r}} \approx \nabla_{\rm{a}}$. OPAL opacity tables are used at high temperature.

For each metallicity 125 evolutionary sequences with logarithmically distributed pre-main sequence masses from 0.08$\,\rm{M_\odot}$ to 150$\,\rm{M_\odot}$ were generated. The 16 metallicities considered in this work are \{0.00001, 0.00003, 0.0001, 0.0003, 0.001, 0.002, 0.003, 0.004, 0.005, 0.006, 0.008, 0.01, 0.02, 0.03, 0.04, 0.05\}. The STARS code only has opacity tables for these metallicities. Relative abundances of metals were taken from \cite{andersgrevse}. The evolutionary sequences with metallicities below 0.002 and initial masses below 0.3$\,\rm{M_\odot}$ or over 70$\,\rm{M_\odot}$ encounter numerical instabilities and often break down either during H or He ignition. 

The stars were run from the top of the Hayashi track on the pre-mainsequence until they broke down. The pre-main-sequence evolution is important for forming representative composition profiles at the start of the main-sequence evolution. When the models break down convective regions, particularly in the envelopes, become numerically unstable and fluctuate substantially in radius and mass. 

The STARS code is not able to self-consistently ignite degenerate material. Models below 2.25$\,\rm{M_\odot}$ were run to the tip of the RGB then break down when the star undergoes degenerate He-ignition at the helium flash. Models between 2.25$\, \rm{M_\odot}$ and 7$\, \rm{M_\odot}$ ignite He non-degenerately. These models either crash during the first thermal pulse of the AGB or when the stars attempt to degenerately ignite carbon. The models over 7$\, \rm{M_\odot}$ ignite He, evolve along the AGB, ignite carbon and then break down during oxygen ignition.

\subsection{Structure Dependent Tidal Calculations}
\begin{figure*}
    \centering
        \includegraphics[width = \textwidth]{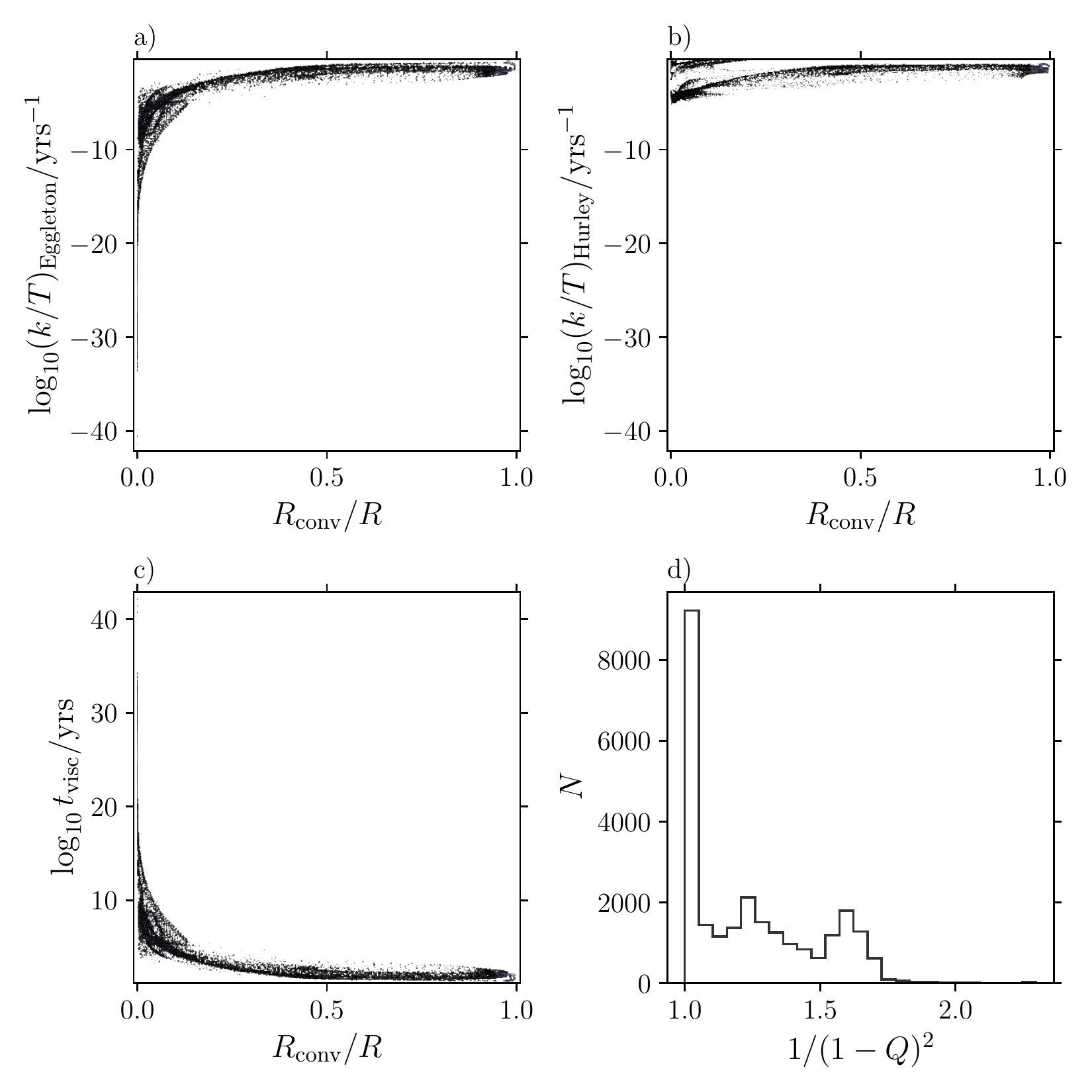}
    \caption{Various computed tidal quantities for the the stellar models with metallicity $z = 0.02$. Subplot is a) he structure dependent $(k/T)_{\rm{c}}$ calculated according to \cite{eggtides} and \cite{hut} as a function of fractional convective radius, b) is the structure dependent $(k/T)_{\rm{c}}$ calculated according to \cite{bse} and \cite{rasio} as a function of fractional convective radius, c) is the contribution to $(k/T)_{\rm{c}}$ from $t_{\rm{visc}}$ according to \cite{eggtides}and d) is a histogram showing the distribution of the stellar structure constant $Q$ calculated according to \cite{eggtides}.}
    \label{fig:tvisccalc}
\end{figure*}

For each calculated evolutionary stellar model the structure dependent terms for tidal evolution, $Q$ and $t_{\rm{visc}}$ can be calculated using the formalism laid out above. If $t_{\rm{visc}}$ and $Q$ are known for a given evolutionary stellar model, $(k/T)_{\rm{c}}$ is also known and vise versa. Fig. \ref{fig:tvisccalc} shows the computed $(k/T)_{\rm{c}}$ according to \cite{eggtides} and\cite{hut} with the $t_{\rm{visc}}$ and $Q$ and the approximate $(k/T)_{\rm{c}}$ according to \cite{bse}. Both estimates of $(k/T)_{\rm{c}}$ show a general trend of increasing $(k/T)_{\rm{c}}$ for increasing $(r_{\rm{conv}}/R)$. Both estimates also show that, for the same $(r_{\rm{conv}}/R)$, the $(k/T)_{\rm{c}}$ shows considerable scatter and can vary by some orders of magnitude. Comparison between \cite{eggtides} and \cite{bse} show that the two estimates of $(k/T)_{\rm{c}}$ differ by many orders of magnitude at small $(r_{\rm{conv}}/R)$. In regions of small $(r_{\rm{conv}}/R)$, \cite{bse} overestimate $(k/T)_{\rm{c}}$ by many orders of magnitude and thus also overestimate the strength of tidal dissipation. The distinct population of \cite{bse}'s $(k/T)_{\rm{c}}$ estimates between $10^{-3}$ and 0 at $(r_{\rm{conv}}/R) < 0.45$ correspond to the convective regions in the cores of the stars. 

The contribution to $(k/T)_{\rm{c}}$ from $t_{\rm{visc}}$ and $Q$ can be seen in Figs. \ref{fig:tvisccalc}c and \ref{fig:tvisccalc}d. Because $Q$ is typically small, $1/(1-Q)^2$ is between 1 and 2.5 with a preference for values close to unity. Hence, the stellar structure constant $Q$ makes only a small contribution to the dissipation strength of the tides. However, $t_{\rm{visc}}$ varies by many orders of magnitude and so is the dominant contribution to $(k/T)_{\rm{c}}$. At small $(r_{\rm{conv}}/R)$ $t_{\rm{visc}}$ shows considerable scatter with estimates varying by up to 30 orders of magnitude.

\subsection{Comparison With Approximations}
\begin{figure*}
    \centering
        \includegraphics[width = \textwidth]{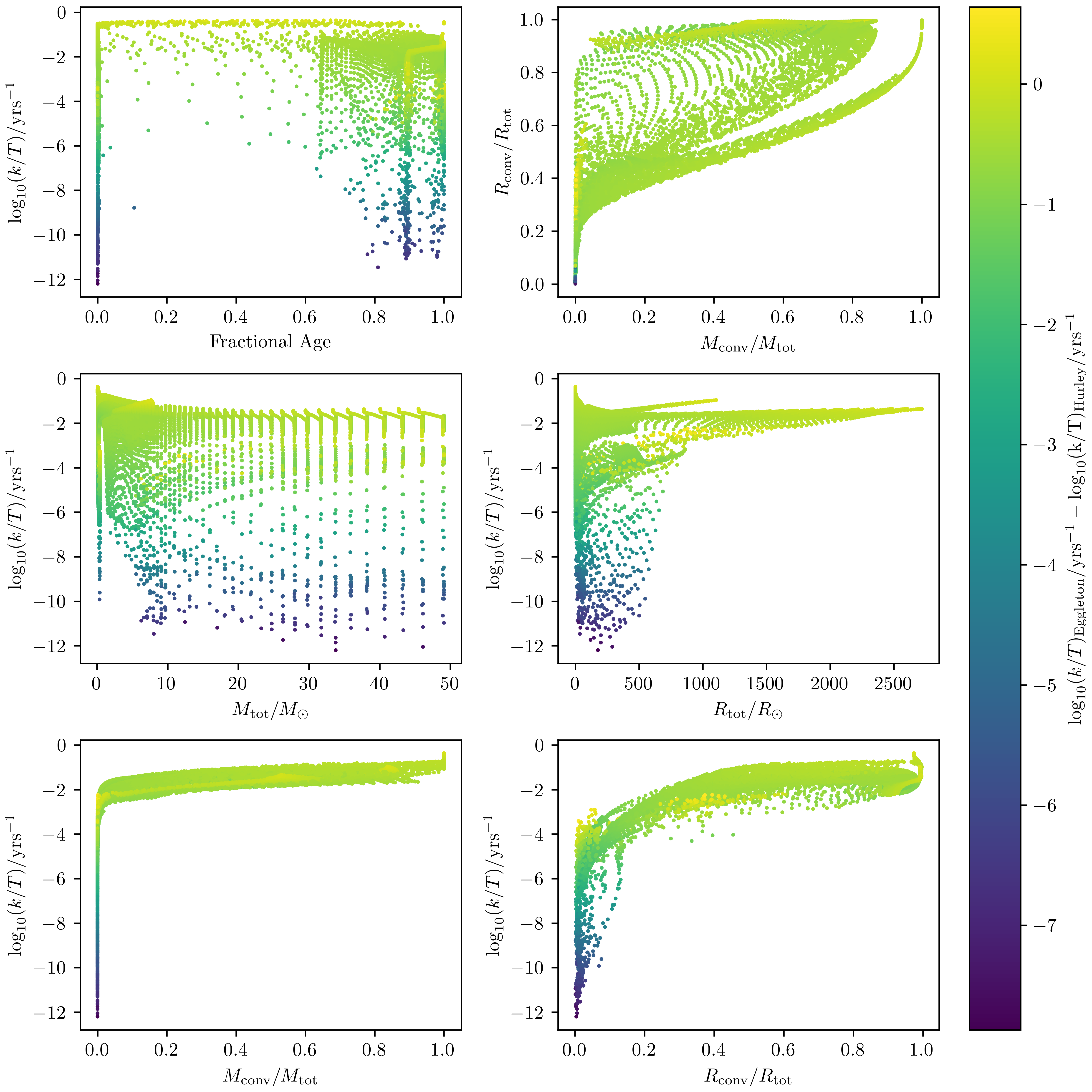}
    \caption{The plotted $(k/T)_{\rm{c}}$ is that of the detailed calculation for the envelope regions for all the stellar models calculated with $z=0.02$. The colorbar shows the difference in $(k/T)_{\rm{c}}$ when calculated using the \cite{bse} prescription (Approx) and when calculated using the formalism of \cite{eggtides} (Calc) then using equation \ref{eq:equivalence}. Top left: $(k/T)_{\rm{c}}$ as a function of age, top right: fractional convective radius as a function of fractional convective mass, middle left: $(k/T)_{\rm{c}}$ as a function of total mass, $(k/T)_{\rm{c}}$ as a function of total radius, bottom left: $(k/T)_{\rm{c}}$ as a function of fractional convective mass, bottom right: $(k/T)_{\rm{c}}$ as a function of fractional convective radius.}
    \label{fig:ktcdiscrepancy}
\end{figure*}

\begin{figure*}
    \centering
        \includegraphics[width = \textwidth]{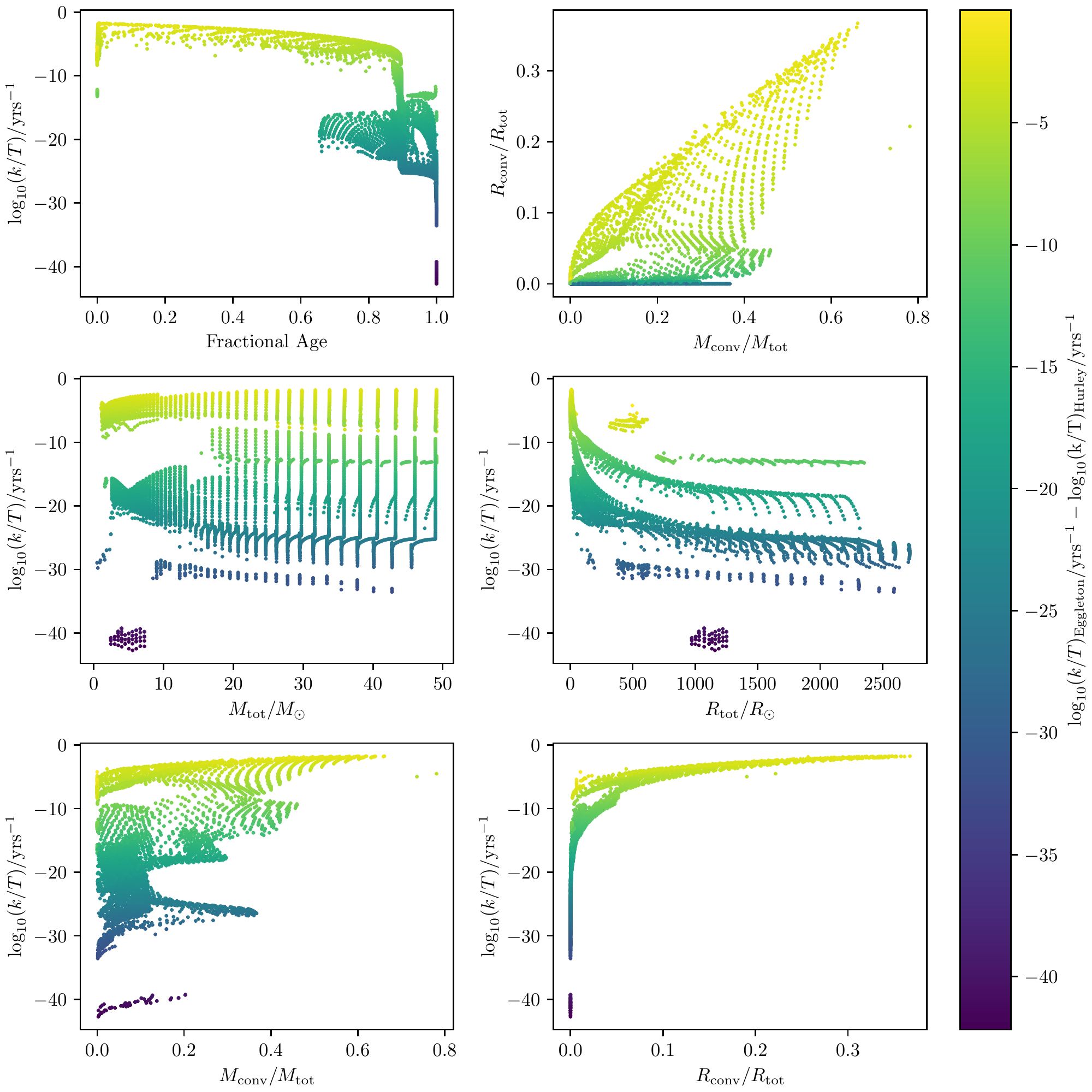}
    \caption{As Fig. \ref{fig:ktcdiscrepancy} but for the convective core regions. }
    \label{fig:ktcdiscrepancycore}
\end{figure*}

 Figs. \ref{fig:ktcdiscrepancy} and \ref{fig:ktcdiscrepancycore} compare the respective envelope and core $(k/T)_{\rm{c}}$ as functions of various stellar parameters. In this work, convective cores refer to convective regions which have an inner boundary at the centre of the star and convective envelopes refer to convective regions with an outer boundary at the surface of the star. The color bar in each plot refers to the discrepancy between \cite{eggtides} and \cite{bse}, defined as $\log_{10} (k/T)_{\rm{Eggleton}} - \log_{10}(k/T)_{\rm{Hurley}}$. 
 
 Fig. \ref{fig:ktcdiscrepancy} reveals that \cite{bse} systematically overestimate $(k/T)_{\rm{c}}$ in envelope regions by close to an order of magnitude and that, as $(k/T)_{\rm{c}}$ decreases, the discrepancy between the two estimates increases by up to 8 orders or magnitude. The $(k/T)_{\rm{c}}$ works well for main-sequence stars but breaks down at parts of the pre-main sequence and the later evolutionary phases. Fractional age is used mostly for visual clarity in the plots because the low-mass stars have lifetimes many orders of magnitude longer than the high-mass stars. Models with masses between $0.8\,\rm{M_\odot}$ and $2.5\,\rm{M_\odot}$ are best approximated. However the lower mass and higher mass models all show orders of magnitudes of difference when $(k/T)_{\rm{c}}$ becomes small. The stars with radii greater than $500\, R_\odot$ have large $(k/T)_{\rm{c}}$ so the approximation works to within an order of magnitude. At the smallest $(m_{\rm{conv}}/M)$ $(k/T)_{\rm{c}}$ rapidly decreases to small values and the approximation totally breaks down. A similar trend is seen for small  $(r_{\rm{conv}}/R)$ although it is less pronounced. 

Fig. \ref{fig:ktcdiscrepancycore} shows that the approximation breaks down almost completely in the convective core regions with the tidal efficiency being overestimated by over 40 orders of magnitude in some cases. The late evolutionary phases are least well captured. Unlike those with convective envelopes, stars with masses between $2.5 \, \rm{M_\odot}$ and $8\, \rm{M_\odot}$ have the largest discrepancies when comparing \cite{bse} and \cite{eggtides}. As with the envelopes, small $(m_{\rm{conv}}/M)$ and $(r_{\rm{conv}}/R)$ show the largest discrepancies. Unlike the envelopes, the core $(k/T)_{\rm{c}}$ rapidly plummets for small $(r_{\rm{conv}}/R)$ and more steadily decreases for small  $(m_{\rm{conv}}/M)$. 

In the core regions a large amount of mass is contained in a small radius whereas in the envelopes a small amount of mass is contained in a large radius. The substantial difference in the density of the core and envelope regions explains the differing behavior for small $(r_{\rm{conv}}/R)$ and $(m_{\rm{conv}}/M)$ and motivates forming separate prescriptions for the convective core regions and convective envelope regions. 

\section{Approximating the Viscous Time}
Figs \ref{fig:ktcdiscrepancy} and \ref{fig:ktcdiscrepancycore} demonstrate that, of the parameters available in BSE, $(k/T)_{\rm{c}}$ is most dependent on the fractional convective mass and radius. We fit a power law of the form 
\begin{equation}
    \bigg( \frac{k}{T}\bigg)_{\rm{c}} = \bigg(\frac{R_{\rm{conv}}}{R}\bigg)^a\bigg(\frac{M_{\rm{conv}}}{M}\bigg)^b \frac{c}{t_{\rm{conv}}}
\end{equation}
for the cores and envelopes for each metallicity set individually. The power law is of a similar form to \cite{bse}'s approximation so is easily implementable. However the inclusion of the radius term allows us to capture the behaviour of $(k/T)_{\rm{c}}$ more accurately. Here we use 
\begin{equation}
        t_{\rm{conv}} = 0.4311 \bigg[ \frac{3M_{\rm{conv}} R_{\rm{conv}}^2)}{L} \bigg]^{1/3}\,\rm{yr},
        \label{eqtconv}
    \end{equation}
so to be able to use the same equation for the core and envelope regions. Comparison of the convective turnover time obtained from mixing length theory \citep{tglob}
\begin{equation}
    t_{\rm{conv}} = \int_{r_{\rm{b}}}^{R}\frac{dr}{w},
\end{equation}
where $r_{\rm{b}}$ is the radius co-ordinate at the base of the convective zone, $R$ is the radius of the star and $w$ is the mixing velocity and that from Eq. \ref{eqtconv} shows that there is a scatter of a factor of 3. Thus $(k/T)_{\rm{c}}$ is fit with this term included to correct for these discrepancies.

\begin{figure*}
    \centering
        \includegraphics[width = \textwidth]{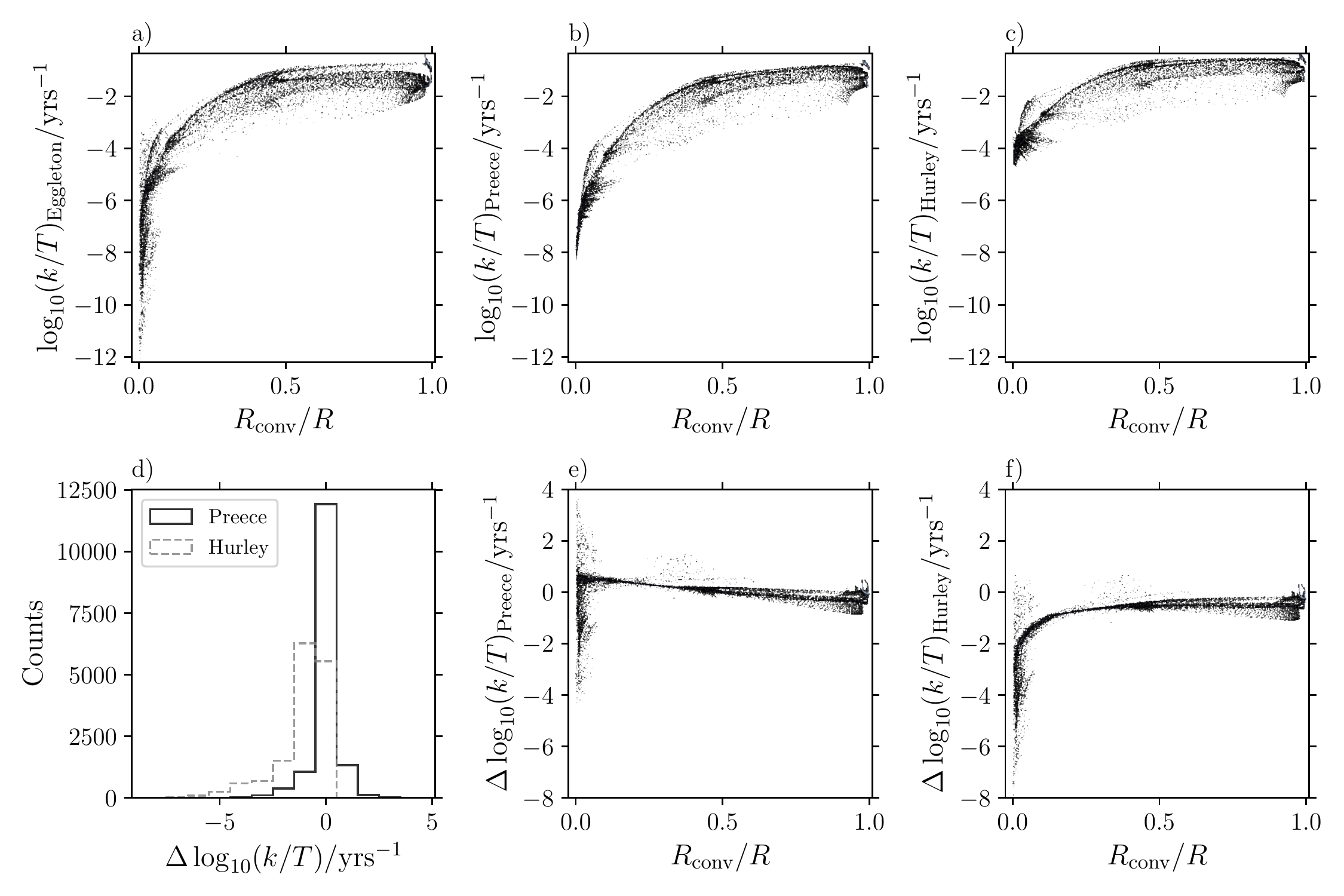}
    \caption{The calculated $(k/T)_{\rm{c}}$ for all convective envelope regions found in the stellar models with $z=0.02$ for a) the formalism of \cite{eggtides}, b) the fit obtained in this work, c) the existing prescription in \cite{bse}. Subplot d) is a histogram of the residuals of $\log_{10}(k/T)_{\rm{c}}$ for our prescription and the \cite{bse} prescription when compared to the detailed calculation, e) are the residuals for our prescription and f) are the residuals for the \cite{bse} prescription.}
    \label{fig:ktcfitresid}
\end{figure*}

 Fig. \ref{fig:ktcfitresid} shows the envelope $(k/T)_{\rm{c}}$ from \cite{eggtides}, the best fit from this work, and \cite{bse}. The individual residuals between our prescription and the \cite{bse} prescription are displayed as a histogram of the residuals from the two prescriptions for $z=0.02$. \cite{bse}'s approximation systematically overestimates $(k/T)_{\rm{c}}$ in all $(m_{\rm{conv}}/M)$ and overestimates $(k/T)_{\rm{c}}$ by many orders of magnitude for fractional convective masses approaching 0. The new fit corrects the systematic offset but still does not capture the full behaviour when fractional convective masses approach 0. When the convective mass approaches 0 the viscous time-scale often exceeds the Hubble time and thus tides are very ineffective. The new fit at least captures that the viscous time becomes very large, so that $(k/T)_{\rm{c}}$ is very small and the tidal evolution is negligible. The histogram of the residuals of the two prescriptions highlights that the new fit is predominantly accurate to within a factor of a few. 

\subsection{Metallicity Dependence}
\begin{figure*}
    \centering
        \includegraphics[width = \textwidth]{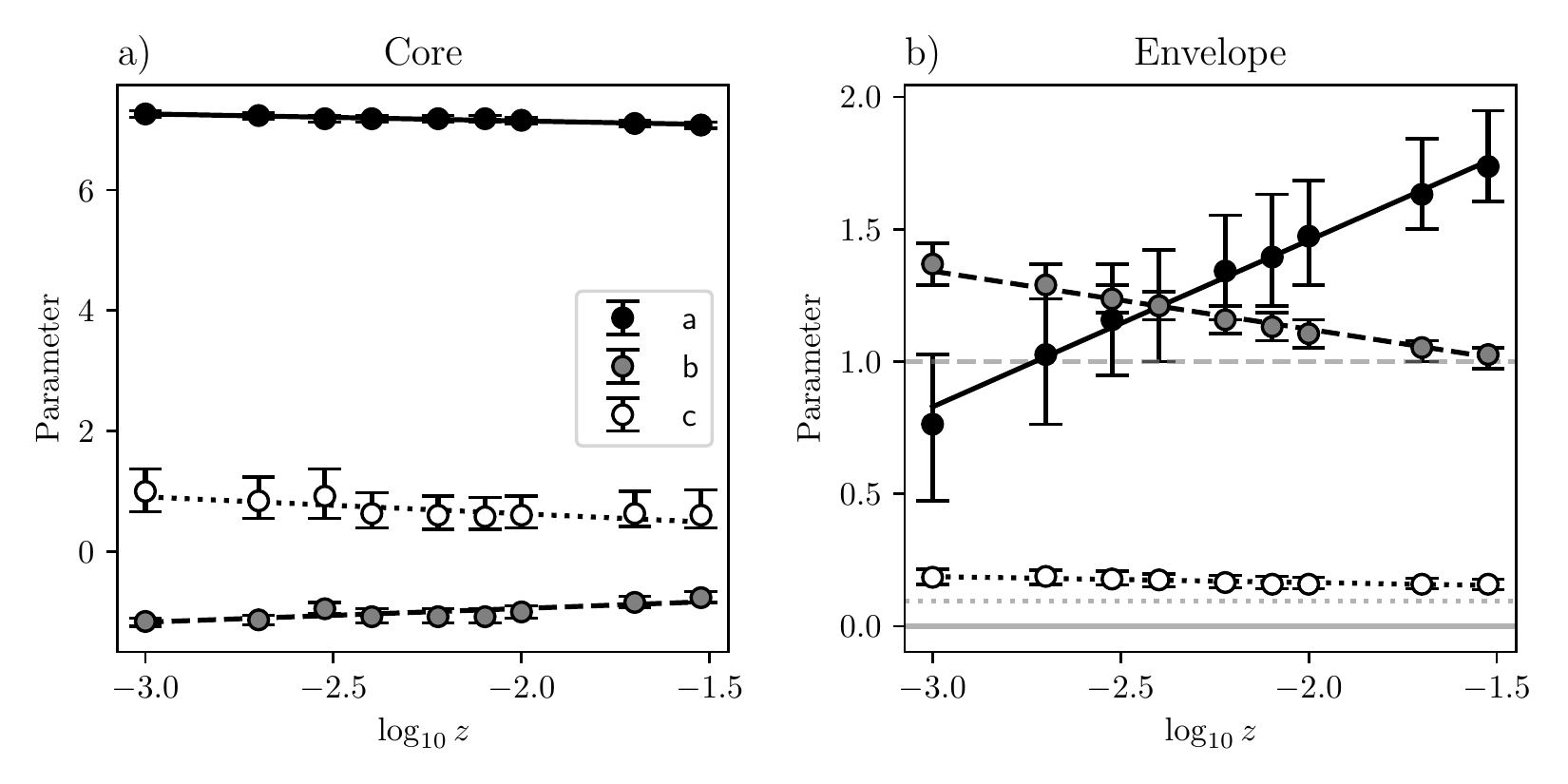}
    \caption{The metallicity dependence of the parameters $a$, $b$ and $c$ from the power law fits for a) the convective core regions and b) the convective envelope regions. The grey lines in the b) are the values of $a$, $b$ and $c$ used in \cite{bse}.}
    \label{fig:abcz}
\end{figure*}

Each metallicity set is fitted separately to obtain an individual $a$, $b$ and $c$ with the procedure described in the above section. Fig. \ref{fig:abcz} shows the obtained parameters. In log-linear space clear trends can be seen in the values of $a$, $b$ and $c$ for differing metallicities. To capture this metallicity dependence, a log-linear straight line fit is carried out for each parameter in the core and envelope such that $a(z)$, $b(z)$ and $c(z)$ are then given by
\begin{equation}
    a(z)=a_1\log_{10}z+a_2,
\end{equation}

\begin{equation}
    b(z)=b_1\log_{10}z+b_2 
\end{equation}
and
\begin{equation}
    c(z)=c_1\log_{10}z+c_2, 
\end{equation}
and the final prescription is given by

\begin{equation}
\bigg( \frac{k}{T}\bigg)_{\rm{c}} = \bigg(\frac{R_{\rm{conv}}}{R}\bigg)^{a(z)}\bigg(\frac{M_{\rm{conv}}}{M}\bigg)^{b(z)} \frac{c(z)}{t_{\rm{conv}}}.
\end{equation} 
 The fits are only carried out for $-3.0 < \log_{10}z<-1.5$, or $0.001 < z < 0.03$, where $z$ is the metallicity. Outside the displayed range the straight line dependence breaks down. Metallicities below $z=0.001$ refer to extremely young population III stars so we consider the range of our fits to be relevant to most existing stellar observations.  
 
 Table \ref{tab:prescriptiomn} shows the obtained values of $a(z)$, $b(z)$ and $c(z)$ for both the viscous time and $(k/T)_{\rm{c}}$. We fit both quantities because $(k/T)_{\rm{c}}$ is the parameter used in BSE and takes into account the contribution of both the viscous time and $1/(1-Q)^2$. The viscous time relates more obviously to the structure of the star, particularly the estimates of the turbulent convective viscosity.

\begin{deluxetable*}{ccccc}[b!]
\tablecaption{Obtained values of $a$, $b$ and $c$ including metallicity dependence for both the viscous time and $(k/T)_{\rm{c}}$ \label{tab:prescriptiomn}}
\tablecolumns{6}
\tablenum{1}
\tablewidth{0pt}
\tablehead{
\colhead{Quantity} &
\colhead{Region} &
\colhead{Parameter $\chi$} & 
\colhead{$\chi_1$} &
\colhead{$\chi_2$} 
}
\startdata
$(k/T)_{\rm{c}}$ & Core & $a(z)$ & -0.12 $\pm$ 0.01 & 6.91 $\pm$ 0.03\\
$(k/T)_{\rm{c}}$ & Core & $b(z)$ &  0.23 $\pm$ 0.04& -0.5 $\pm$ 0.1\\
$(k/T)_{\rm{c}}$ & Core & $c(z)$ & -0.28 $\pm$ 0.08 & 0.07 $\pm$ 0.02 \\ \hline
$(k/T)_{\rm{c}}$ & Envelope & $a(z)$ & 0.63$\pm$0.02 & 2.72 $\pm$ 0.05\\
$(k/T)_{\rm{c}}$ & Envelope & $b(z)$ & -0.219$\pm$0.009 & 0.68 $\pm$ 0.02\\
$(k/T)_{\rm{c}}$ & Envelope & $c(z)$ & -0.023 $\pm$ 0.004 & 0.220 $\pm$ 0.009\\ \hline
$t_{\rm{visc}}$ & Core & $a(z)$ & 0.10 $\pm$ 0.01 & -6.96 $\pm$ 0.03\\
$t_{\rm{visc}}$ & Core & $b(z)$ & -0.24 $\pm$ 0.04& 0.44 $\pm$ 0.1\\
$t_{\rm{visc}}$ & Core & $c(z)$ & -0.09 $\pm$ 0.08 & 0.35 $\pm$ 0.02 \\ \hline
$t_{\rm{visc}}$ & Envelope & $a(z)$ & -0.62$\pm$0.02 & -2.75 $\pm$ 0.05\\
$t_{\rm{visc}}$ & Envelope & $b(z)$ & 0.23$\pm$0.01 & -0.61 $\pm$ 0.02\\
$t_{\rm{visc}}$ & Envelope & $c(z)$ & 0.018 $\pm$ 0.004 & 0.53 $\pm$ 0.02\\
\enddata
\tablecomments{The metallicity dependence of the parameters $a$, $b$, and $c$ fit for both the viscous time and $(k/T)_{\rm{c}}$. Straight lines in log-linear space are fit to the parameters for the metallicity dependence such that $\chi(z) =  \chi_1 \log_{10}z+\chi_2$.}
\end{deluxetable*}

\section{Implementation into BSE}
To evaluate the effect of our updated prescription for tidal coupling on binary evolution we implement our results into BSE. We then run the updated version of BSE and the original version with the same initial conditions and compare the results to see whether the new tidal prescription has a statistically significant effect on the resulting systems. 
\subsection{Initial Conditions}
\begin{figure*}
    \centering
        \includegraphics[width = \textwidth]{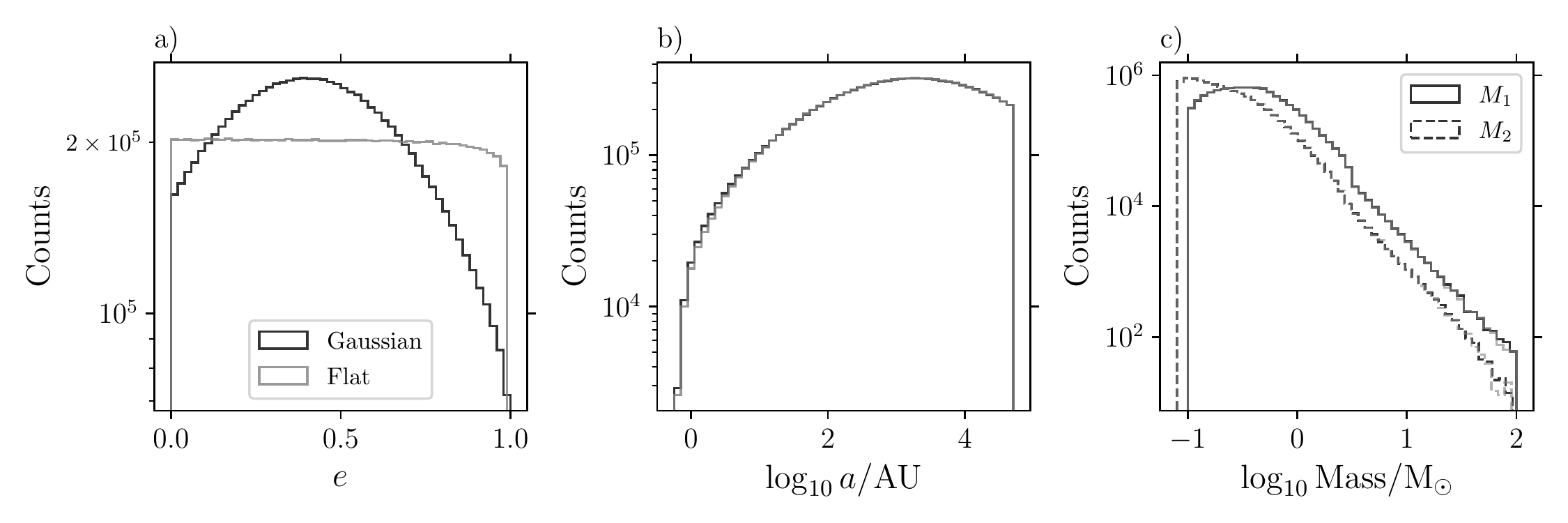}
    \caption{A histogram of the Monte-Carlo sampled initial conditions used for the population synthesis for a) the eccentricity, $e$ b) the semi-major axis, $\log a$. c) the masses of the primary and secondary. Of the four sampled eccentricity distributions two are displayed - a flat and Gaussian distribution. The initial conditions for both are shown in the plot with the Gaussian distribution show in black and the flat in gray.}
    \label{fig:icall}
\end{figure*}

The initial conditions for the population synthesis are generated with Monte-Carlo sampling as by \cite{tedi} but for the binary case. The primary masses are selected with a 3 component broken power law relation in the range of 0.08 to 100 $\rm{M_\odot}$ \citep{kroupa, kroupaii}. The secondary masses are uniformly selected based on the mass ratio of the primary to the secondary with the additional condition that the secondary mass does not exceed the primary mass. The semi-major axis follows a log-normal distribution if the primary mass is below $3\,\rm{M_\odot}$ \citep{lowma} and is flat in $\log a$ for higher masses \citep{highma} with the additional criteria that neither star can have a radius exceeding its Roche radius, and that $a < 10^3 \,\rm{AU}$. The upper limit on the semi-major axis is introduced because, at larger separations than this, the tidal interactions do not play a significant role and so are not of interest here. Each system is evolved for 14$\,$Gyrs. Fig. \ref{fig:icall} shows the initial distributions of $M_1$, $M_2$, $a$ and $e$ for the population synthesis for both a Gaussian and flat eccentricity distribution. The flat eccentricity distribution gives slightly fewer systems at small separations and slightly fewer high-mass objects from the requirement that the system must not be filling its Roche lobe at the start of the evolution. 

\subsubsection{Eccentricity Distributions}

We consider four separate eccentricity distributions for this work. We generate data sets with and without our new tidal prescription with four distinct eccentricity distributions and solar metallicity. Each data set contains $10^7$ systems. We use a thermal eccentricity distribution \citep{jeanse}, a Gaussian distribution of eccentricities with a mean and standard deviation of 0.4, a flat eccentricity distribution and a set with all systems circularized. The Gaussian and flat distributions are considered to be more representative of the observed binary population \citep{multiplicity1}. In the majority of the following analysis we show only the results for the Gaussian distributed eccentricity data sets for brevity. The differences in the results obtained for the different eccentricity distributions are small.

\subsection{Results}
In our population synthesis analysis we focus on the final state of the systems. We consider the final orbital parameters of the surviving binaries, the final stellar type, the spin orbit synchronization and the Ia supernova rate.

\subsubsection{Analysis of An Individual System}

\begin{figure*}
    \centering
        \includegraphics[width = \textwidth]{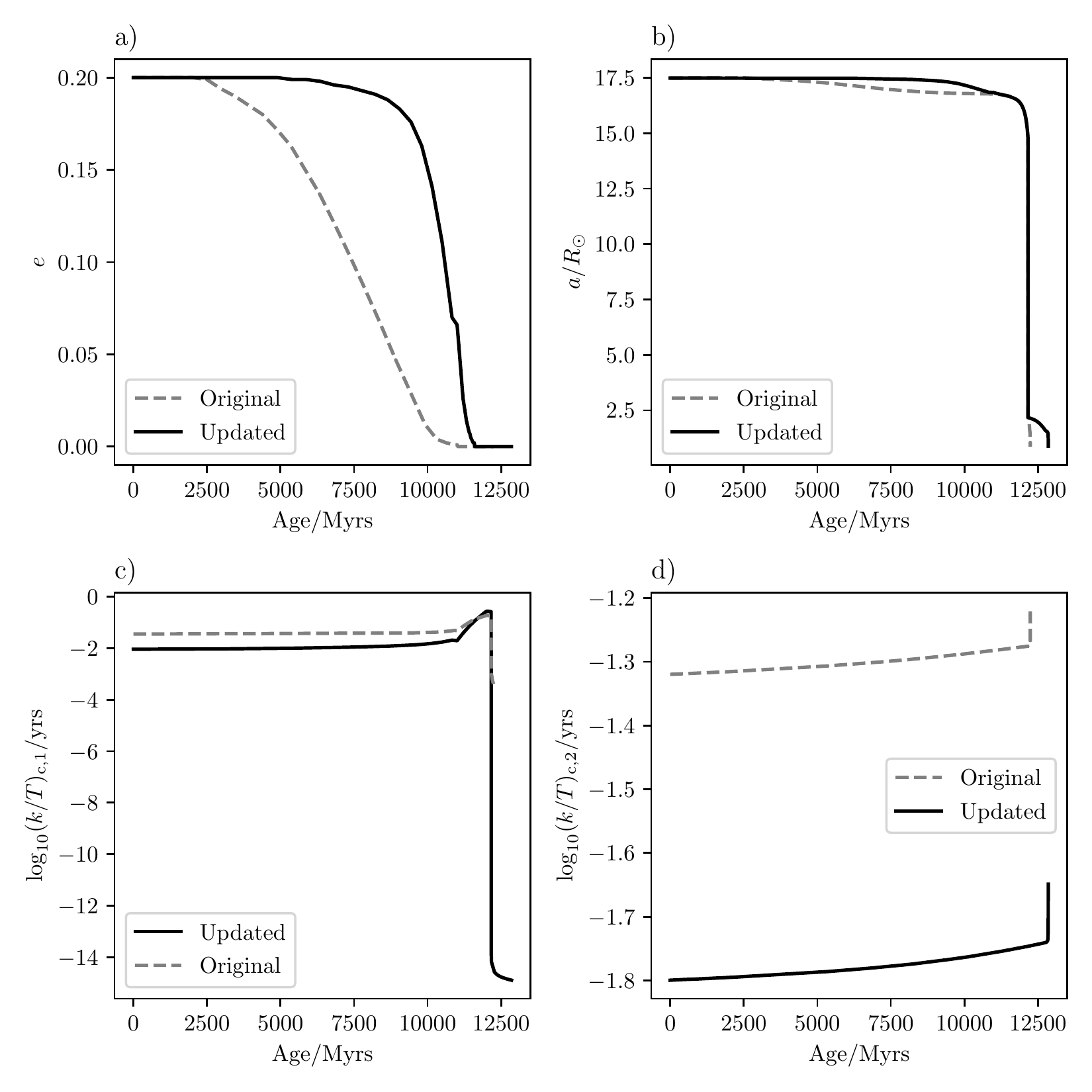}
    \caption{The evolution of a binary with initial masses of $1\,\rm{M_{\odot}}$ and $0.8\,\rm{M_{\odot}}$, an initial orbital period of 6.32 days and an initial eccentricity of 0.2. Subplot a) is the eccentricity evolution is, b) is the semi-major axis evolution, c) and d) give the $(k/T)_{\rm{c}}$ from the convective region for primary and secondary star respectively. }
    \label{fig:detailed}
\end{figure*}

To assess the impact of our tides prescription we follow the evolution of a binary with initial masses of $1\,\rm{M_{\odot}}$ and $0.8\,\rm{M_{\odot}}$, an initial orbital period of 6.32 days and an initial eccentricity of 0.2. The results are shown in Fig. \ref{fig:detailed}. With both tides prescriptions the more massive primary finishes its main-sequence evolution after 11000 Myrs and evolves to become Herzsprung gap star. At 11590 Myrs the primary becomes a red giant branch star, then the envelope is removed and the primary becomes a He-WD. Next, the WD accretes some mass from the main-sequence secondary then the two merge to form an RGB star. The newly formed single RGB star evolves to become a CO-WD. The change in the tides prescription affects both the circularization and merger time. With the original prescription the binary is almost fully circularized by the end of the main-sequence however with the new prescription the binary only circularizes at the start of the RGB. Further, the updated prescription increases the time taken for the merging of the WD and MS. 

\subsubsection{Final Orbital Parameters}
The final orbital parameters of all systems for the Gaussian  distributed eccentricity data set are shown in Fig \ref{fig:finalagauss}. The other results from the other eccentricity distributions are in Appendix 1. BSE gives an eccentricity of $-1$ to  binaries which have been broken apart by supernova kicks or systems which have merged or Ia supernovae which have destroyed one of the stars in the system. The number of systems with $e=-1$ decreases by 14\% with the updated prescription for the Gaussian distributed eccentricities. The distribution of primary masses is somewhat altered at low masses. The mass plots include stars with $e=-1$ but exclude all mass-less remnants.

The final semi-major axis distribution is not significantly affected by the new tides prescription. Minor changes to the semi-major axis distributions can be seen at small $a$ for all the eccentricity distributions. Note that the systems with $e=-1$ and systems including a mass-less remnant are not included in the semi-major axis plot. Marginally more systems with small semi-major axis are found with the updated tides prescription owing to the reduced merging rate. The systems with sufficiently small initial separations to experience tidal interactions also undergo at least one phase common-envelope evolution, if the stars are sufficiently massive that one or both objects evolves off the main sequence within the Hubble time. Additionally, the total number of circularized systems is not strongly affected. A small excess of non-circularized systems with small semi-major axis can be seen suggesting the critical semi-major axis for tidal circularization is increased. Common-envelope evolution is far more efficient than tidal interactions at shrinking and circularizing the orbit so the effect of the tides often cannot be seen in the final system. The final state of the data sets with an initially circular eccentricity are least affected by the change in tides.

\begin{figure*}
    \centering
        \includegraphics[width = \textwidth]{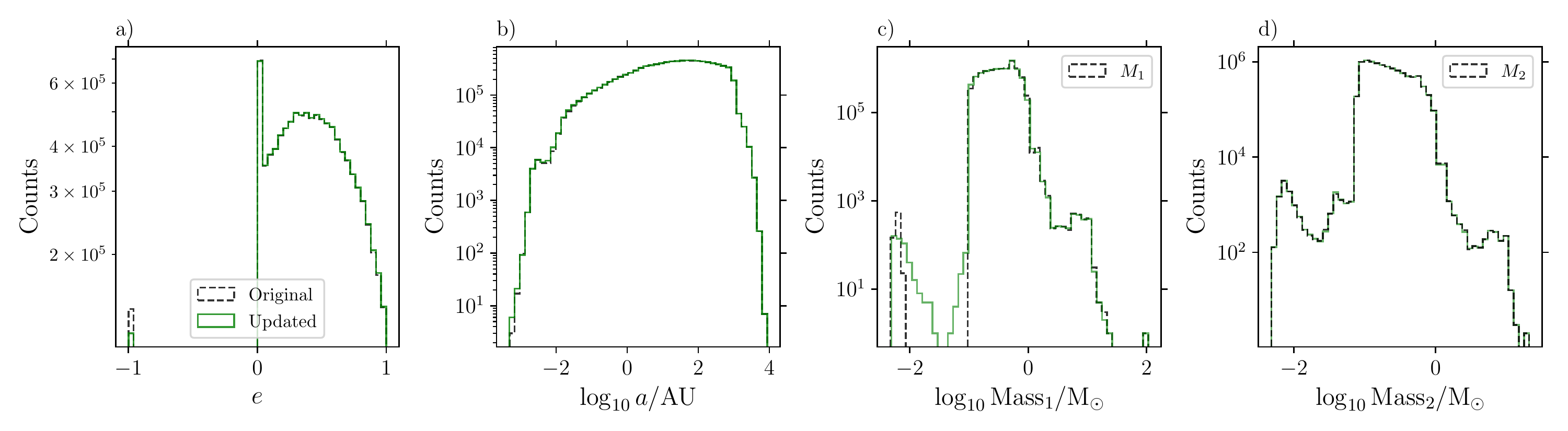}
    \caption{The final orbital parameters of the data set with an initially Gaussian eccentricity distribution. Subplot a) shows the eccentricity at the end of the evolution of all all the systems, b) is the final distribution of the semi-major axis for all the surviving binary systems, c) is the mass of all the surviving primaries and d) show the mass distribution of the secondaries. }
    \label{fig:finalagauss}
\end{figure*}

\subsubsection{Spin-Orbit Synchronization}
\begin{figure}
    \centering
        \includegraphics[width = \columnwidth]{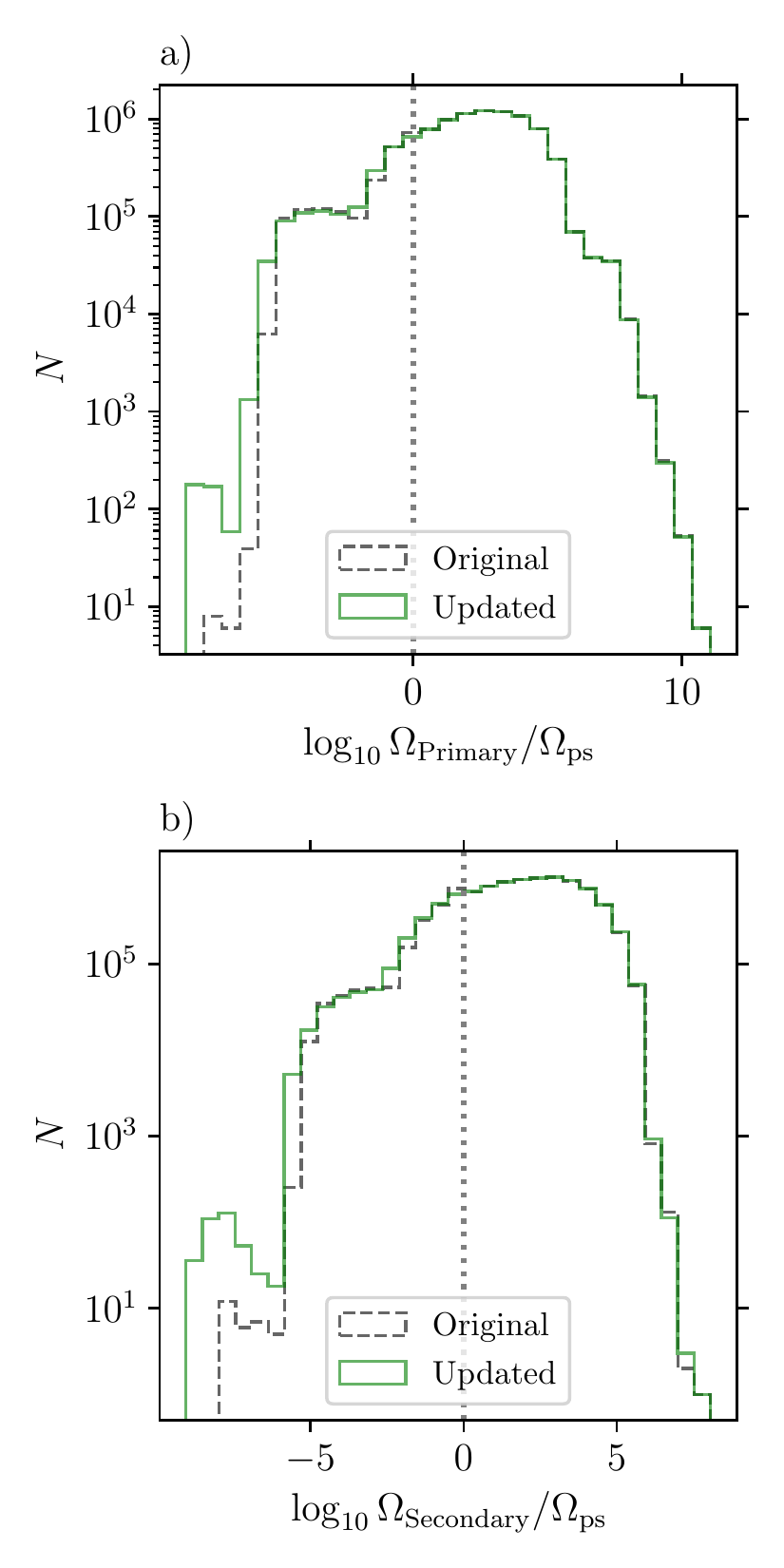}
    \caption{The ratio of the rotation rate to the corresponding pseudo-synchronous rotation rate for the surviving binaries with initially Gaussian distributed eccentricities. Subplot a) are the primary stars and b) are the secondary stars.}
    \label{fig:sync}
\end{figure}

Spin-orbit synchronization is a useful measure of the effectiveness of the tides because common envelopes do not substantially change the spin rate of the stars in BSE. Tidal synchronization in circular binaries is achieved when the spin period of the star is the same as the orbital period. In eccentric orbits tidal interactions are strongest at periastron and weakest at apastron. Pseudo-synchronization is achieved when $\dot{\Omega}=0$. This approximately occurs when the spin frequency is comparable to the orbital motion at periastron \citep{hut,pseudo}. The pseudo-synchronization frequency can be calculated as
\begin{equation}
    \Omega_{\rm{ps}}= \frac{1 + (15/2)e^2 + (45/8)e^4 + (5/16)e^6 }{(1+3e^2 + (3/8)e^4)(1-e^2)^{3/2}} \omega.
\end{equation}
For circular orbits $\Omega_{\rm{ps}} = \omega$.
Fig. \ref{fig:sync} shows the ratio of the spin frequency to the pseudo-synchronization frequency for the surviving binaries for the data sets with Gaussian distributed eccentricities. Unless otherwise specified, BSE assumes the empirical relation for initial spins of \cite{intialspins}. All systems with $\log_{10}(\Omega/\Omega_{\rm{ps}})$ close to 0 are considered to be synchronized. Systems with small $(\Omega/\Omega_{\rm{ps}})$ are rotating sub-synchronously and systems with large $(\Omega/\Omega_{\rm{ps}})$ are rotating super-synchronously. The majority of super-synchronous objects are in wide binaries which have long orbital periods and spin periods determined by single star evolution. Some CO WDs in close binaries which have accreted some matter from their companion are also rapid rotators. We find a similar number of systems achieve spin-orbit synchronization with both tidal prescriptions. However we observe a clear excess of sub-synchronously rotating stars far from tidal synchronization with our new tidal prescription.

\subsubsection{Orbital Circularization}

\begin{figure}
    \centering
        \includegraphics[width = \columnwidth]{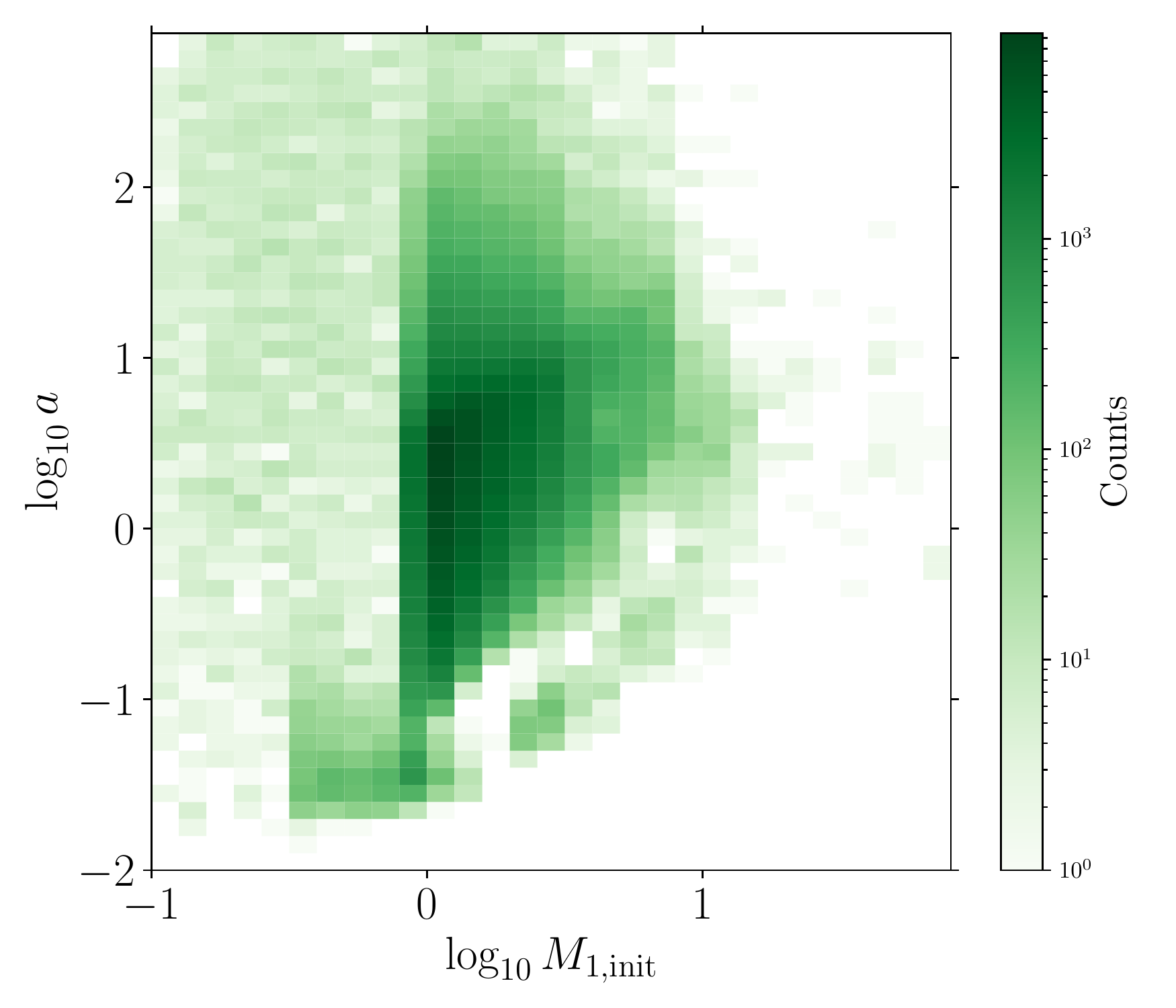}
    \caption{A 2D histogram showing the initial distribution of primary mass and semi-major axis for the circularized systems for the new tides prescription in the Gaussian distributed data set.}
    \label{fig:initalcirc}
\end{figure}

\begin{figure}
    \centering
        \includegraphics[width = \columnwidth]{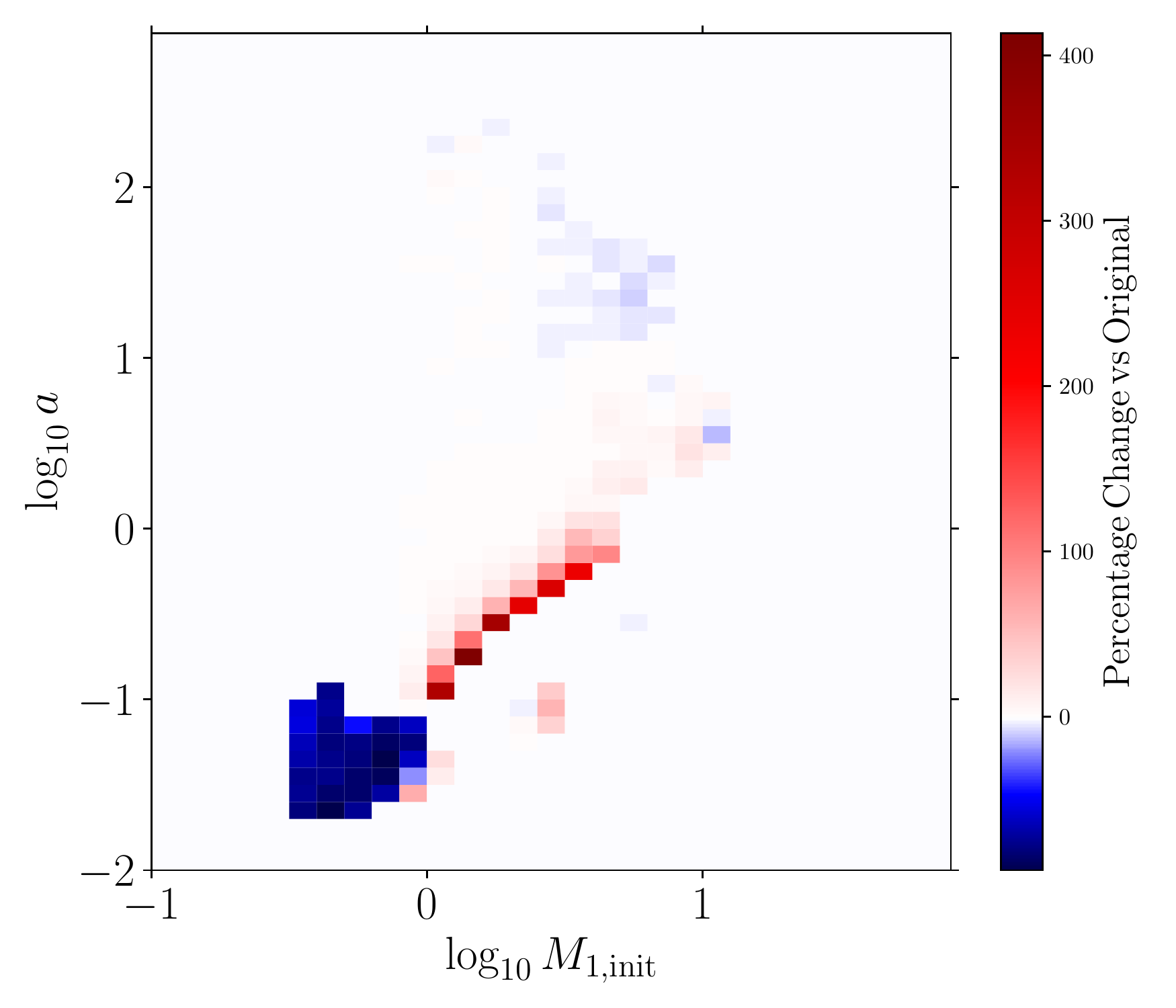}
    \caption{A 2D histogram showing the change in the initial distribution of primary mass and semi-major axis of the circularized systems when comparing the original and updated tides prescription in the Gaussian distributed data set.}
    \label{fig:intialcircdiff}
\end{figure}

Fig. \ref{fig:initalcirc} shows the 2D parameter space of the initial primary mass and initial semi-major axis of the circularized systems with the updated tides prescription for the Gaussian distributed data set. Fig. \ref{fig:intialcircdiff} shows the difference in the initial conditions of the circularized systems when comparing the two tides prescriptions. The number of circularized systems with low masses decreases. The number of merged systems involving initially low-mass stars also decreases. These low-mass primaries predominantly do not have time to evolve off the main sequence. As seen when examining the individual system in Sec. 5.2.1, the new prescription causes the binaries to circularize less efficiently on the main-sequence. For the stars with small semi-major axis and masses between 1$\,\rm{M_\odot}$ and 8$\,\rm{M_\odot}$ the tides are sufficiently efficient to circularize the systems during the post-main sequence evolution. However fewer binaries merge so there are more circularized systems. The number of circularized systems with masses above 1$\,\rm{M_\odot}$ and initial semi-major axis greater than 1 AU decreases by a few percent.

\subsubsection{Orbital Parameters at Onset of Common-Envelope}
\begin{figure*}
    \centering
        \includegraphics[width = \textwidth]{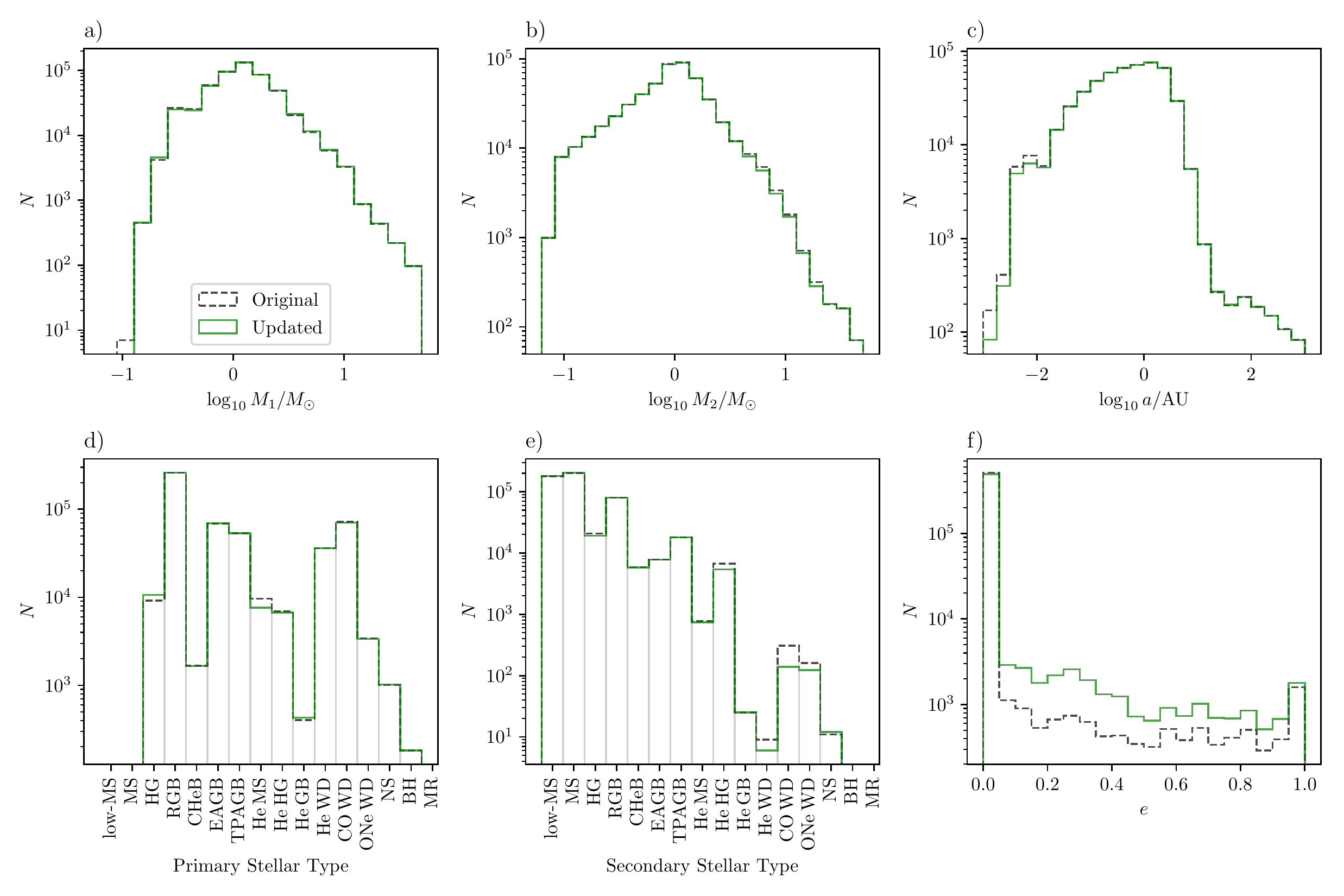}
    \caption{Various parameters at the start of the mass-transfer phase which leads to the common envelope evolution. The distributions are a) the masses of the primaries, b) the masses of the secondaries, c) the semi-major axes, d) the stellar types of the primaries, e) the stellar types of the secondaries and f) the eccentricities. The solid green line represents the results with the updated tides prescription, the dashed black line shows the original BSE result. The data set with initially flat distributed eccentricities are displayed.}
    \label{fig:ceeonset}
\end{figure*}

The distributions of the orbital parameters at the onset of the mass-transfer episode which results in common-envelope evolution can be seen in Fig. \ref{fig:ceeonset}. In this figure we consider the flat distributed data set so as to assess the impact on eccentricity more clearly. The Gaussian distributed data set had too few system at high eccentricity to make meaningful statistical inferences. The primary and secondary masses at the start of the common-envelope evolution are unaffected by the change in tides prescription. The new tides prescription decreases the incidence of common-envelope at low semi-major axis. Fewer common envelope events occur in systems with CO WD secondaries with the updated prescription. The majority of systems are circularized at the start of the mass transfer episode leading to common envelope however the number of systems with $e>0$ increases by close to an order of magnitude with the new tides prescription. BSE allows for non-zero eccentricities both at the onset of common-envelope and in the resulting system. As described in \cite{mse}, the post-common envelope eccentricity of the system is
\begin{equation}
    e^2_{\rm{fin}} = 1 - E_{\rm{orb,fin}}\bigg(\frac{1  - e^2_{\rm{init}}}{E_{\rm{orb,init}}}\bigg),
\end{equation} 
where $E_{\rm{orb,init}}$ and $E_{\rm{orb,fin}}$ are the respective initial and final orbital energies and $e_{\rm{init}}$ is the eccentricity at the onset of common envelope. If $E_{\rm{orb,fin}} > E_{\rm{orb,init}}$ the post-CE system is circularized.

\subsubsection{Merging}
The frequency of merging decreases by 16.5\% overall when considering the Gaussian distributed binaries. Of these affected merges, the majority are low-mass MS + low-mass MS binaries. The He-WD + MS/low-mass MS which form CO-WD's are the next most substantially impacted. Many other evolutionary branches are impacted to a small degree.

\subsubsection{Stellar Type}

\begin{figure}
    \centering
        \includegraphics[width = \columnwidth]{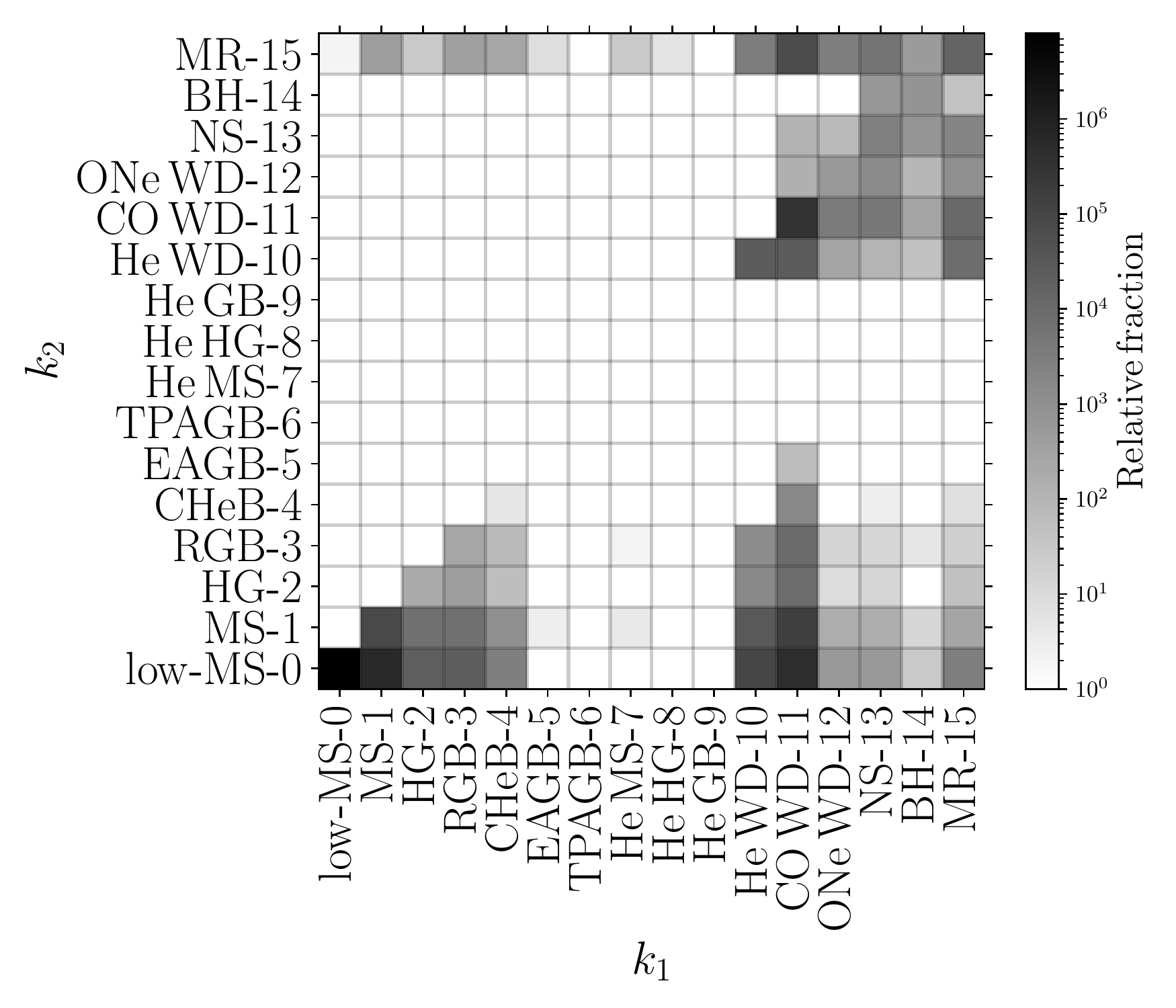}
    \caption{A 2D histogram showing the final type of star 1 and star 2 after 14 Gyrs with the new tides prescription for the data set with initially Gaussian distributed eccentricities. Note that stellar type 15 refers to massless remnants which are formed either by Ia supernova or merges.}
    \label{fig:chess1}
\end{figure}

\begin{figure}
    \centering
        \includegraphics[width = \columnwidth]{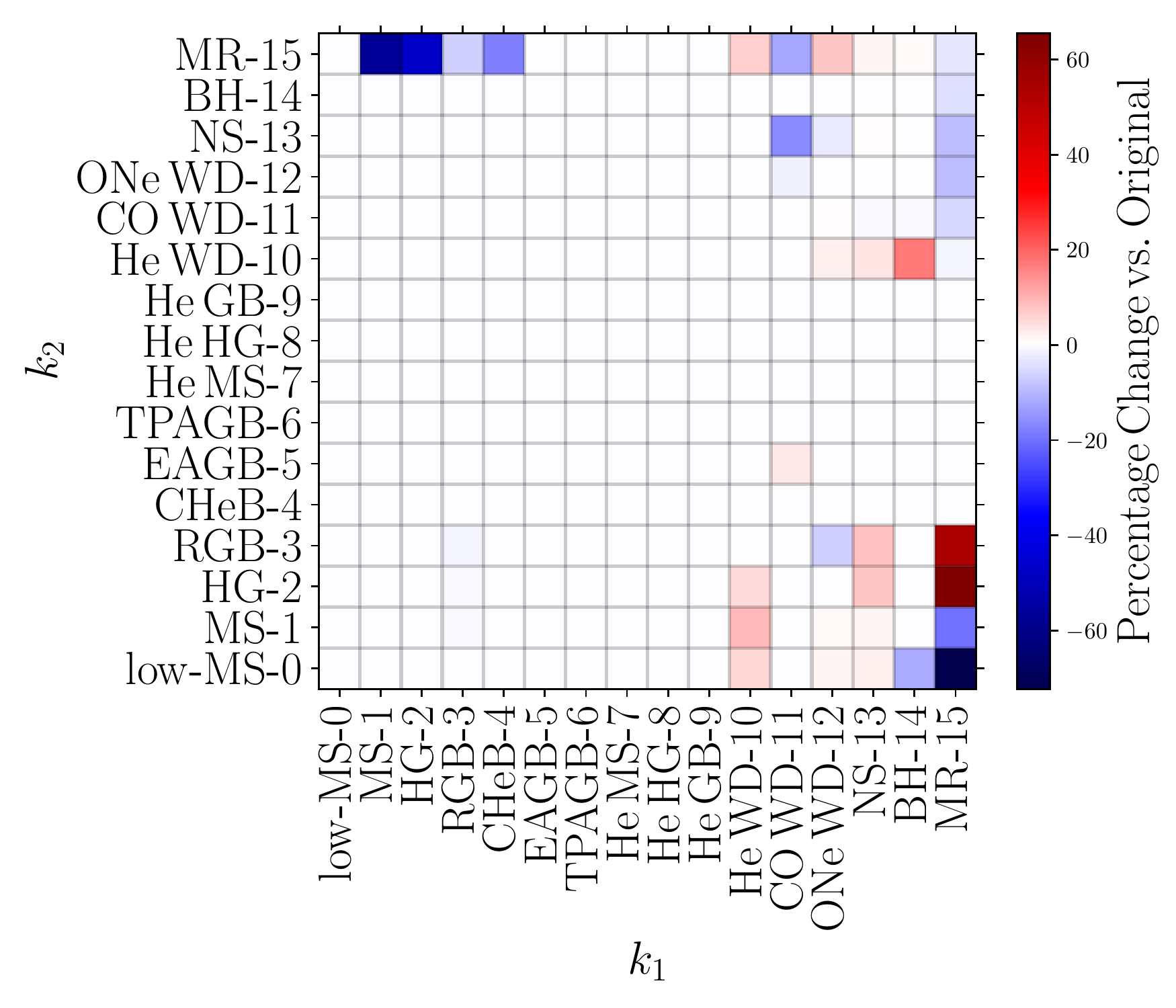}
    \caption{A 2D histogram showing the percentage change in systems when comparing the \cite{bse} tides prescription to the updated prescription for the data set with initially Gaussian distributed eccentricities. In the blue regions the original \cite{bse} scheme produces more of these systems, in the red region the updated prescription produces more}
    \label{fig:chess2}
\end{figure}

The final stellar type of the objects with our new prescription is shown in Fig. \ref{fig:chess1} for the Gaussian distributed eccentricity set. Fig. \ref{fig:chess2} shows the percentage change in the number of systems with each stellar type when compared to the unmodified BSE. The resulting stellar type of the binaries is seen to be influenced by our tides prescription. The single-star population, with so called massless remnants as the companions, is the most significantly effected owing to decreased merging and thermonuclear supernovae (such as supernovae of Type Ia) rate. BSE designates the disrupted stars in merges and post Ia supernovae CO-WDs as massless remnants. The decrease in single low-mass main-sequence and main-sequence stars can be attributed to a decrease in early merging, primarily between two main-sequence objects. The increase in binary He-WDs and decrease in single CO-WDs is predominantly due to a decrease in the number of He-WD/COWD + low-mass MS/MS merges which subsequently form CO-WDs. The decrease in systems where both stars are destroyed or where the primary becomes a single HeWD or ONe-WD relates to an altered Ia supernova rate.

When looking at the final binary parameters by individual stellar types it can be seen that the final semi-major axis distribution of the main-sequence stars is most altered. This is likely due to the fact that these systems have either not experienced common-envelope evolution, if both stars are main-sequence, or undergone one phase of common-envelope evolution, if the companion is a more massive evolved star. Common-envelope evolution tends to diminish the relative effect of the tides, since the former typically affect the orbit much more significantly. 

\subsubsection{Occurrence of thermonuclear Supernovae}
As an observationally somewhat tractable application of our new prescription, we discuss the impact on production rates of progenitors of thermonuclear supernovae. We are using the SNe Ia channels as originally employed by \cite{bse}. 
The number of thermonuclear supernovae detected in the population syntheses are shown in Table \ref{tab:sne}. The overall occurrence of thermonuclear supernovae is reduced by 3.8\% for the Gaussian distributed eccentricities. Several evolutionary channels are proposed produce different thermonuclear supernovae, of which supernovae of type Ia are a subtype, of which the dominant channel is still a matter of debate \citep[see][for a recent review]{R2020}. The channels with the CO-WD as a primary or secondary are distinguished in Table \ref{tab:sne} because although the explosion mechanism is the same, the formation channels are different. In our selection of initial conditions the primary star is always initially more massive. The double degenerate systems with a CO-WD primary and He-WD secondary, the initially more massive star becomes a CO-WD and the initially less massive star evolves more slowly to form a He-WD. Some mass from the primary is accreted on to the secondary such that it can evolve into a He-WD within the Hubble time. The systems with a He-WD primary and CO-WD secondary are Algol systems. Mass transfer early in the binary evolution cause the initially less massive secondary star to gain enough mass from the primary to become the more massive star in the system. In this case the initially less massive secondary evolves to become a CO-WD and the initially more massive secondary becomes a He-WD.  According to our simulations, the dominant thermonuclear supernova channel is an accretion induced detonation of a CO-WD with a He-WD. In BSE He rich material from the He-WD is accreted on to the CO-WD via Roche lobe overflow. Once $0.15\,\rm{M_\odot}$ of material have been accreted, the CO-WD explodes. The BSE limit of $0.15\,\rm{M_\odot}$ is motivated by 1D stellar evolution models \citep[and similar prescriptions are used in other population synthesis studies, such as that by][]{WJH2013}. However this has been challenged in more recent simulations \citep[][]{YL2004a,WK2011}. This channel, otherwise known as the double detonation scenario, suggests that sufficient material accreted quiescently, from the donor, will, under compressional heating, ignite. The following He-detonation then propagates into the CO-core, leading to a thermonuclear SN \citep{N1980P,N1982a,N1982b}. 
The prescription for this event currently implemented in BSE does not take the known influence of the mass transfer rate and fluctuations thereof \citep{WK2011} into account, instead assuming a SN to occur once sufficient material has been transferred. Depending on the mass transfer rate, systems of this type may instead lead to a subsonic ignition (at high mass transfer rates) of the helium. This would likely result in a massive helium nova or a so-called .Ia \citep{BSWN2007,KHG2014} or fast, faint flashes \citep[e.g.][]{PTY2014}.
This channel is not substantially affected by the changed tides prescription, possibly a result of the the SN prescription. However all other subdominant channels are altered statistically significantly. The rate of merging of binary CO-WDs is reduced by 16\%. The single degenerate evolution channels (where He-MS+CO-WD channels are potentially associated with SNe Iax and CO-WD+MS and CO-WD+HG potentially with archetypal SNe Ia) as a whole are reduced by 12.8\%.  
These systems have only undergone one phase of common-envelope evolution and have non-degenerate stellar companions which are influenced by the tides. The rare scenario of CO-WD + MS/HG to HG Ia supernovae rate is the only single degenerate channel which sees an increase in the rate. 

We find that the systems with high initial eccentricities are less likely to undergo a thermonuclear supernova, as defined by \cite{bse}. The data set with the thermal eccentricity distribution, and thus the most high eccentricity systems, is the most affected by the changed tides prescription. Owing to the reduced early merging rate, more systems with smaller initial separations and lower mass primaries undergo a thermonuclear supernova with our updated prescription. The double degenerate CO-WD + CO-WD channel for the data set with initially circular orbits is not affected by the new tides prescription. 
This discussion strongly suggests the conclusion that detailed knowledge of tidal interaction is crucial in the study of transients depending on close binary interaction, such as SNe Ia. 

\begin{deluxetable*}{cccc|ccc|ccc|ccc|ccc}[b!]
\rotate
\tablecaption{Ia Supernovae occurrence by channel \label{tab:sne}}

\tablecolumns{7}
\tablenum{2}
\tablewidth{0pt}
\tablehead{
\colhead{$k_{1,\rm{SNe}}$} &
\colhead{$k_{2,\rm{SNe}}$} &
\colhead{$k_{1,\rm{Fin}}$} &
\colhead{$k_{2,\rm{Fin}}$} &
\colhead{Tot 1} &
\colhead{Tot 2} &
\colhead{\% Diff.} &
\colhead{Tot 1} &
\colhead{Tot 2} &
\colhead{\% Diff.} &
\colhead{Tot 1} &
\colhead{Tot 2} &
\colhead{\% Diff.} &
\colhead{Tot 1} &
\colhead{Tot 2} &
\colhead{\% Diff.} 
}
\startdata
$e$ distribution & & & &  & Gauss. &  &  & Flat &  &  & Therm. &  &  & Circ. & \\ \hline
CO-WD & He-WD & - & He-WD & 9206 & 9078 & -1.4 & 9451 & 9347 & -1.1 & 10666 & 11010 & +3 & 7958 & 8015 & +0.7\\
CO-WD & CO-WD & - & - & 1220 & 942  & -16 & 1045 & 893 & -15 & 801 & 619 & -23 & 1294 & 1298 & + 0.3\\
He-WD & CO-WD & He-WD & - & 941  & 1105 & +17 & 840 & 988 &+18 & 639 & 752 & +17 & 1095 & 1302  & +19\\ \hline
Double Degen. & & & & & & -1.3\% & & & -1\% &&& +2.3\% &&& +2.6\%  \\\hline
CO-WD & MS/HG/RGB/ & - & CO-WD & 2135 & 1925 & -9 & 2119 & 1935 & -8 & 2104 & 1991 & -5 & 2107 & 2005 & -5\\
 & HeMS/TPAGB & & & & &  &&&&&&&&&\\
CO-WD & He-MS & - & NS & 254 & 237 & -6 & 238 & 224 & -6 & 162 & 147 & -9 & 379 & 384 & +1\\
He-MS & CO-WD & CO-WD & - & 234 & 124 & -47 & 244 & 130 & -47 & 172 & 74 & -57 & 282 & 101 & -64 \\
CO-WD & MS & - & MS & 99 & 75 & -24 & 100 & 89 & - 11 & 118 & 68 & -4 & 79& 66 & -16\\
CO-WD & MS/HG & - & HG & 28 & 42 & +50 & 33 & 45 & +36 & 34 & 42 & + 29 & 18 & 40 & +125\\
CO-WD & MS/HG & - & RGB & 16 & 14 & -12 & 18 & 21 & +16 & 34 & 29 & -14 & 17 & 17 & 0\\ \hline
Single Degen. & & & & & & -12.8\%  &&& -10.5\% &&& -10.4\% &&& -9.2\%\\ \hline
Total & & & & & & -3.6\% &&& -2.8\%  &&& -4.6\% &&& -1.0\%\\ \hline
\enddata
\tablecomments{$k_1$ refers to the stellar type of the initially more massive primary star, $k_2$ refers to the stellar type of the less massive secondary. $k_{1,\rm{Sne}}$ and $k_{2,\rm{Sne}}$ at SNe refer to the stellar type at the time of supernova and the final $k_{1,\rm{Fin}}$ and $k_{2,\rm{Fin}}$ refer to the stellar type at the end of the simulation. The Tot 1 refers to the total number of systems found for each channel using the \cite{bse} tides prescription and the Tot 2 refers to the number of systems from our updated prescription. The \% Diff field compares the number of supernova for the two tides prescriptions for the same eccentricity distribution.}
\end{deluxetable*}

\section{Discussion}
As can be seen from the results obtained, the updated prescription has a statistically significant effect on the outcome of the population synthesis. It is important that relatively small changes in the tidal prescription can, for the closest binary systems, have a tangible effect on the overall stellar evolution. Our prescription dramatically decreases the merging rate and modifies the thermonuclear supernova rate. The orbital evolution of the binaries is likely influenced more strongly than the final state of the surviving binaries suggests. Common-envelope evolution substantially shrinks the orbit and circularizes the binary more efficiently than the tides, and also erases any previous tidal locking. The tides are likely to play more of a role in the pre-common envelope evolution and thus effect the common envelope initial conditions.

The main area of uncertainty in the theoretical formalisms presented here lies in the estimation of the convective viscosity. As suggested by \cite{eggtides}, the viscous time-scale is a parameter that may be amenable to observation. The improved accuracy of the prescription means that population synthesis results can be compared to observations to provide insight into convective viscosity.

The theory presented here assumes solid body rotation of the star. Whilst this may be a good approximation for stars that are fully convective, for stars which have substantial convective and radiative regions it is more likely that the stars experience differential rotation. The convective region spins up quickly then, depending on the efficiency of the angular momentum transport, spin up of the radiative regions occurs on a longer time-scale. Small surface convective regions are likely to spin up very rapidly because these regions are the most radially distended but contain relatively little mass. 

Evolved stars, particularly in the high-mass region, typically have multiple convective regions in the middle of the star owing to shell burning. These regions are not identified in BSE but they can occasionally dominate the tidal evolution.

Whilst there are uncertainties in the theory of \cite{eggtides}, particularly with regard to the convective viscosity, it is still useful to have a prescription that has been rigorously compared with the detailed calculations. The updated prescription presented in this work can be said to reduce the discrepancy between the theory and the approximation and thus used to test classical tidal theory with some confidence. 

\section{Conclusions}
We have presented an updated prescription for tidal dissipation in convective regions via the equilibrium tide suitable for population synthesis codes. The prescription was formed by generating a large grid of detailed stellar models and calculating the necessary quantities to estimate the tidal dissipation as formalised by \cite{eggtides}. Next, power laws were fitted to the envelope regions and core regions separately. Metallicity dependent prescriptions for the equilibrium tide in convective cores and convective envelopes were obtained. 

The updated prescription, complete with metallicity dependence was implemented into BSE. The results of the BSE simulations show that a reduction in the merging rate, with those including main-sequence stars most affected. As a further application, the population synthesis also revealed a 12.8\% decrease in the single degenerate thermonuclear supernova channel and a 16\% decrease in the double degenerate merging of two CO white dwarfs when assuming a Gaussian eccentricity distribution. Double degenerate channels featuring a CO-WD primary and He-WD secondary are not impacted. Owing predominantly to the change in the merging and thermonuclear supernova rate the distribution of stellar types at the end of the population synthesis simulations is altered with an increased number of binaries surviving. We found a comparable number of systems having achieved spin orbit synchronization with both prescriptions but also measure an increased number of sub-synchronously rotating systems far from spin-orbit synchronization with our updated prescription. The number of systems which have some eccentricity at the start of common-envelope evolution remains small but nonetheless increases by an order of magnitude.

The results of this paper show a relatively small change in the tides prescription can have a statistically significant effect on the final stellar type, spin-orbit synchronization and observably traceable events such as the rate of early merging and thermonuclear supernova rates. 


\section*{ACKNOWLEDGEMENTS}
CAT thanks Churchill College for his fellowship. A.S.H. thanks the Max Planck Society for support through a Max Planck Research Group. We thank the anonymous referee for their valuable comments.

\appendix

\section{Final Orbital Parameters for All BSE Data Sets}
\begin{figure*}
    \centering
        \includegraphics[width = \textwidth]{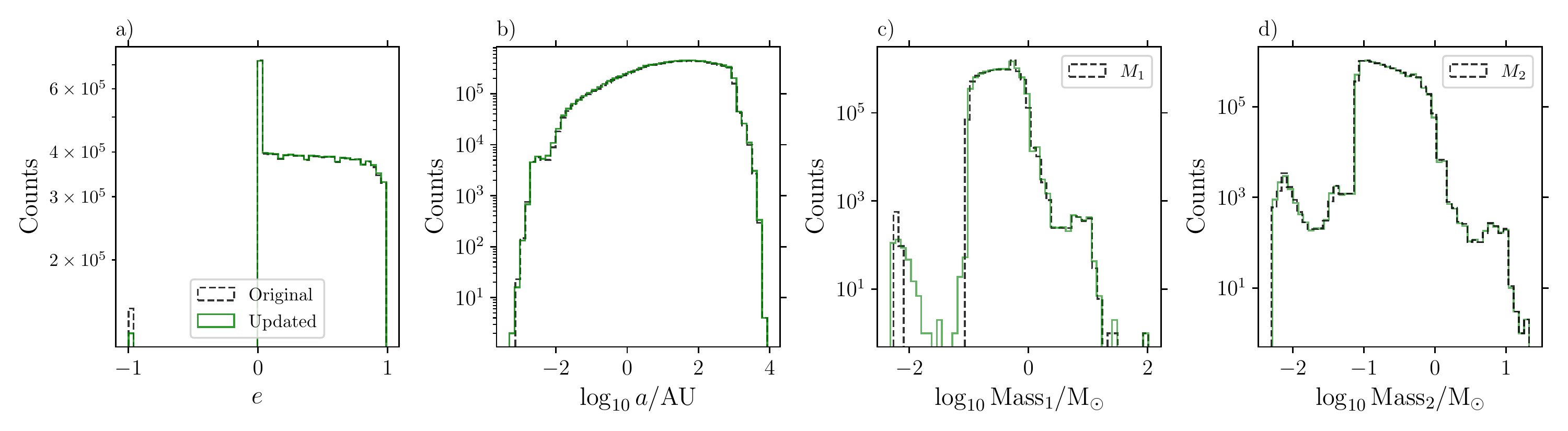}
    \caption{The final orbital parameters of the population synthesis data set with a flat eccentricity distribution.}
    \label{fig:finalaflat}
\end{figure*}

\begin{figure*}
    \centering
         \includegraphics[width = \textwidth]{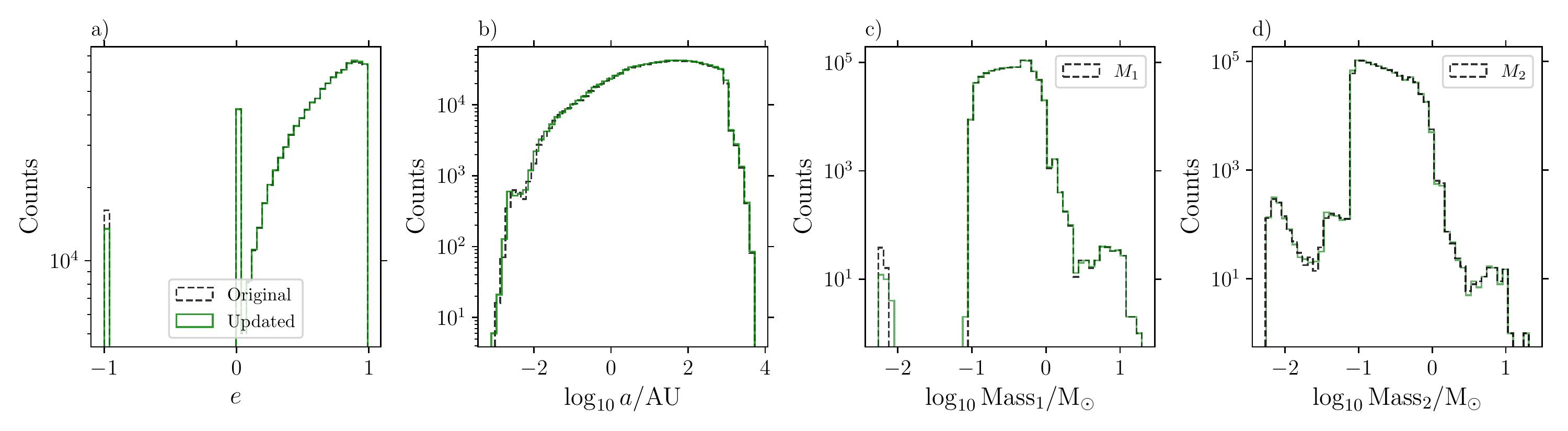}
    \caption{The final orbital parameters of the population synthesis data set with a thermal eccentricity distribution.}
    \label{fig:finalathermal}
\end{figure*}

\begin{figure*}
    \centering
        \includegraphics[width = \textwidth]{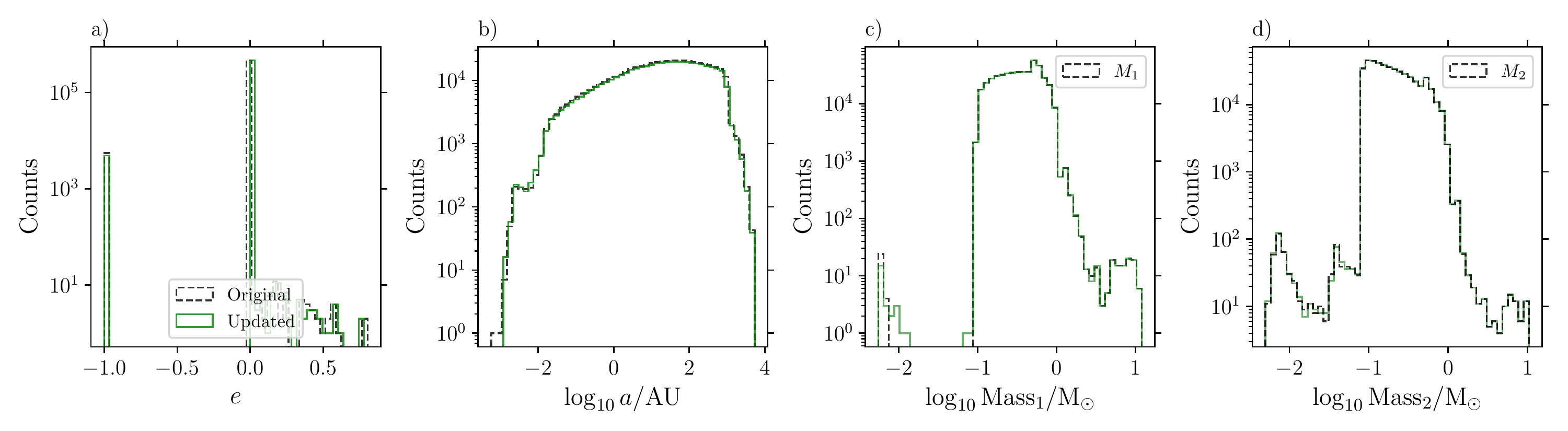}
    \caption{The final orbital parameters of the population synthesis data set with an initially circular eccentricity distribution.}
    \label{fig:finalacirc}
\end{figure*}

For completeness the final orbital parameters for all the population synthesis data sets are shown here. Figs. \ref{fig:finalaflat}, \ref{fig:finalathermal} and \ref{fig:finalacirc} have initial eccentricity distributions which are flat, thermal and circularized respectively.

\bibliography{references}{}

\begin{thebibliography}{}
\expandafter\ifx\csname natexlab\endcsname\relax\def\natexlab#1{#1}\fi
\providecommand{\url}[1]{\href{#1}{#1}}
\providecommand{\dodoi}[1]{doi:~\href{http://doi.org/#1}{\nolinkurl{#1}}}
\providecommand{\doeprint}[1]{\href{http://ascl.net/#1}{\nolinkurl{http://ascl.net/#1}}}
\providecommand{\doarXiv}[1]{\href{https://arxiv.org/abs/#1}{\nolinkurl{https://arxiv.org/abs/#1}}}

\bibitem[{{Alexander}(1973)}]{alexander}
{Alexander}, M.~E. 1973, \apss, 23, 459, \dodoi{10.1007/BF00645172}

\bibitem[{{Anders} \& {Grevesse}(1989)}]{andersgrevse}
{Anders}, E., \& {Grevesse}, N. 1989, \gca, 53, 197,
  \dodoi{10.1016/0016-7037(89)90286-X}

\bibitem[{{Barker} \& {Astoul}(2021)}]{barker}
{Barker}, A.~J., \& {Astoul}, A. A.~V. 2021, \mnras, 506, L69,
  \dodoi{10.1093/mnrasl/slab077}

\bibitem[{{Belczynski} {et~al.}(2016){Belczynski}, {Holz}, {Bulik}, \&
  {O'Shaughnessy}}]{gwsource}
{Belczynski}, K., {Holz}, D.~E., {Bulik}, T., \& {O'Shaughnessy}, R. 2016,
  \nat, 534, 512, \dodoi{10.1038/nature18322}

\bibitem[{{Bildsten} {et~al.}(2007){Bildsten}, {Shen}, {Weinberg}, \&
  {Nelemans}}]{BSWN2007}
{Bildsten}, L., {Shen}, K.~J., {Weinberg}, N.~N., \& {Nelemans}, G. 2007,
  \apjl, 662, L95, \dodoi{10.1086/519489}

\bibitem[{{B{\"o}hm-Vitense}(1958)}]{mlt}
{B{\"o}hm-Vitense}, E. 1958, \zap, 46, 108

\bibitem[{{Claeys} {et~al.}(2014){Claeys}, {Pols}, {Izzard}, {Vink}, \&
  {Verbunt}}]{sne1aii}
{Claeys}, J.~S.~W., {Pols}, O.~R., {Izzard}, R.~G., {Vink}, J., \& {Verbunt},
  F.~W.~M. 2014, \aap, 563, A83, \dodoi{10.1051/0004-6361/201322714}

\bibitem[{{Crawford}(1955)}]{crawfordalgol}
{Crawford}, J.~A. 1955, \apj, 121, 71, \dodoi{10.1086/145965}

\bibitem[{{Darwin}(1879)}]{darwin}
{Darwin}, G.~H. 1879, Philosophical Transactions of the Royal Society of London
  Series I, 170, 1

\bibitem[{{de Mink} \& {Belczynski}(2015)}]{selmabh}
{de Mink}, S.~E., \& {Belczynski}, K. 2015, \apj, 814, 58,
  \dodoi{10.1088/0004-637X/814/1/58}

\bibitem[{{Duch{\^e}ne} \& {Kraus}(2013)}]{multiplicity1}
{Duch{\^e}ne}, G., \& {Kraus}, A. 2013, ARAA, 51, 269,
  \dodoi{10.1146/annurev-astro-081710-102602}

\bibitem[{{Duguid} {et~al.}(2020){Duguid}, {Barker}, \& {Jones}}]{fasttides}
{Duguid}, C.~D., {Barker}, A.~J., \& {Jones}, C.~A. 2020, \mnras, 497, 3400,
  \dodoi{10.1093/mnras/staa2216}

\bibitem[{{Duquennoy} \& {Mayor}(1991)}]{lowma}
{Duquennoy}, A., \& {Mayor}, M. 1991, \aap, 500, 337

\bibitem[{Eggleton(2006)}]{eggzbook}
Eggleton, P. 2006, Evolutionary Processes in Binary and Multiple Stars,
  Cambridge Astrophysics (Cambridge University Press),
  \dodoi{10.1017/CBO9780511536205}

\bibitem[{{Eggleton}(1971)}]{eggz71}
{Eggleton}, P.~P. 1971, \mnras, 151, 351, \dodoi{10.1093/mnras/151.3.351}

\bibitem[{{Eggleton}(1972)}]{starssemiconv}
---. 1972, \mnras, 156, 361, \dodoi{10.1093/mnras/156.3.361}

\bibitem[{{Eggleton}(1983)}]{eggroche}
---. 1983, \apj, 268, 368, \dodoi{10.1086/160960}

\bibitem[{{Eggleton} {et~al.}(1998){Eggleton}, {Kiseleva}, \& {Hut}}]{eggtides}
{Eggleton}, P.~P., {Kiseleva}, L.~G., \& {Hut}, P. 1998, \apj, 499, 853,
  \dodoi{10.1086/305670}

\bibitem[{{Eggleton} \& {Kiseleva-Eggleton}(2001)}]{eggzkltides}
{Eggleton}, P.~P., \& {Kiseleva-Eggleton}, L. 2001, \apj, 562, 1012,
  \dodoi{10.1086/323843}

\bibitem[{{Eggleton} \& {Kisseleva-Eggleton}(2006)}]{eggzkltides2}
{Eggleton}, P.~P., \& {Kisseleva-Eggleton}, L. 2006, \apss, 304, 75,
  \dodoi{10.1007/s10509-006-9078-z}

\bibitem[{{Fabrycky} \& {Tremaine}(2007)}]{fabzkltides}
{Fabrycky}, D., \& {Tremaine}, S. 2007, \apj, 669, 1298, \dodoi{10.1086/521702}

\bibitem[{Fuller \& Lai(2012)}]{fuller}
Fuller, J., \& Lai, D. 2012, Monthly Notices of the Royal Astronomical Society,
  no–no, \dodoi{10.1111/j.1365-2966.2011.20320.x}

\bibitem[{{Goldreich} \& {Nicholson}(1977)}]{chrissuggestion}
{Goldreich}, P., \& {Nicholson}, P.~D. 1977, ICARUS, 30, 301,
  \dodoi{10.1016/0019-1035(77)90163-4}

\bibitem[{{Hamers} {et~al.}(2021{\natexlab{a}}){Hamers}, {Perets}, {Thompson},
  \& {Neunteufel}}]{tedi}
{Hamers}, A.~S., {Perets}, H.~B., {Thompson}, T.~A., \& {Neunteufel}, P.
  2021{\natexlab{a}}, arXiv e-prints, arXiv:2107.13620.
\newblock \doarXiv{2107.13620}

\bibitem[{{Hamers} {et~al.}(2021{\natexlab{b}}){Hamers}, {Rantala},
  {Neunteufel}, {Preece}, \& {Vynatheya}}]{mse}
{Hamers}, A.~S., {Rantala}, A., {Neunteufel}, P., {Preece}, H., \& {Vynatheya},
  P. 2021{\natexlab{b}}, \mnras, 502, 4479, \dodoi{10.1093/mnras/stab287}

\bibitem[{Hoyle(1955)}]{hoyle55}
Hoyle, F. 1955, London, Heinemann Educational, 1970., -1

\bibitem[{{Hurley} {et~al.}(2000){Hurley}, {Pols}, \& {Tout}}]{sse}
{Hurley}, J.~R., {Pols}, O.~R., \& {Tout}, C.~A. 2000, \mnras, 315, 543,
  \dodoi{10.1046/j.1365-8711.2000.03426.x}

\bibitem[{{Hurley} {et~al.}(2002){Hurley}, {Tout}, \& {Pols}}]{bse}
{Hurley}, J.~R., {Tout}, C.~A., \& {Pols}, O.~R. 2002, \mnras, 329, 897,
  \dodoi{10.1046/j.1365-8711.2002.05038.x}

\bibitem[{{Hut}(1981)}]{hut}
{Hut}, P. 1981, \aap, 99, 126

\bibitem[{{Jeans}(1919)}]{jeanse}
{Jeans}, J.~H. 1919, \mnras, 79, 408, \dodoi{10.1093/mnras/79.6.408}

\bibitem[{{Kilic} {et~al.}(2014){Kilic}, {Hermes}, {Gianninas}, {Brown},
  {Heinke}, {Ag{\"u}eros}, {Chote}, {Sullivan}, {Bell}, \& {Harrold}}]{KHG2014}
{Kilic}, M., {Hermes}, J.~J., {Gianninas}, A., {et~al.} 2014, \mnras, 438, L26,
  \dodoi{10.1093/mnrasl/slt151}

\bibitem[{{Kiseleva} {et~al.}(1998){Kiseleva}, {Eggleton}, \&
  {Mikkola}}]{zkltides}
{Kiseleva}, L.~G., {Eggleton}, P.~P., \& {Mikkola}, S. 1998, \mnras, 300, 292,
  \dodoi{10.1046/j.1365-8711.1998.01903.x}

\bibitem[{{Kobulnicky} \& {Fryer}(2007)}]{highma}
{Kobulnicky}, H.~A., \& {Fryer}, C.~L. 2007, \apj, 670, 747,
  \dodoi{10.1086/522073}

\bibitem[{{Kopal}(1959)}]{kopalroche}
{Kopal}, Z. 1959, {Close binary systems}

\bibitem[{{Kozai}(1962)}]{kozai}
{Kozai}, Y. 1962, \aj, 67, 591, \dodoi{10.1086/108790}

\bibitem[{{Kroupa}(2001)}]{kroupa}
{Kroupa}, P. 2001, \mnras, 322, 231, \dodoi{10.1046/j.1365-8711.2001.04022.x}

\bibitem[{{Kroupa}(2002)}]{kroupaii}
---. 2002, Science, 295, 82, \dodoi{10.1126/science.1067524}

\bibitem[{Landin {et~al.}(2010)Landin, Mendes, \& Vaz}]{tglob}
Landin, N.~R., Mendes, L. T.~S., \& Vaz, L. P.~R. 2010, Astronomy and
  Astrophysics, 510, A46, \dodoi{10.1051/0004-6361/200913015}

\bibitem[{{Lang}(1992)}]{intialspins}
{Lang}, K.~R. 1992, {Astrophysical Data I. Planets and Stars.}

\bibitem[{{Lidov}(1962)}]{lidov}
{Lidov}, M.~L. 1962, \planss, 9, 719, \dodoi{10.1016/0032-0633(62)90129-0}

\bibitem[{{Moe} \& {Di Stefano}(2017)}]{multiplicity2}
{Moe}, M., \& {Di Stefano}, R. 2017, ApJ Supplement, 230, 15,
  \dodoi{10.3847/1538-4365/aa6fb6}

\bibitem[{{Nomoto}(1980)}]{N1980P}
{Nomoto}, K. 1980, in Texas Workshop on Type I Supernovae, ed. J.~C. {Wheeler},
  164--181

\bibitem[{{Nomoto}(1982{\natexlab{a}})}]{N1982a}
{Nomoto}, K. 1982{\natexlab{a}}, \apj, 253, 798, \dodoi{10.1086/159682}

\bibitem[{{Nomoto}(1982{\natexlab{b}})}]{N1982b}
---. 1982{\natexlab{b}}, \apj, 257, 780, \dodoi{10.1086/160031}

\bibitem[{{Paczy{\'n}ski}(1971)}]{paczyroche}
{Paczy{\'n}ski}, B. 1971, \araa, 9, 183,
  \dodoi{10.1146/annurev.aa.09.090171.001151}

\bibitem[{{Paczy{\'n}ski} \& {Sienkiewicz}(1972)}]{masstrasfer}
{Paczy{\'n}ski}, B., \& {Sienkiewicz}, R. 1972, \actaa, 22, 73

\bibitem[{{Piersanti} {et~al.}(2014){Piersanti}, {Tornamb{\'e}}, \&
  {Yungelson}}]{PTY2014}
{Piersanti}, L., {Tornamb{\'e}}, A., \& {Yungelson}, L.~R. 2014, \mnras, 445,
  3239, \dodoi{10.1093/mnras/stu1885}

\bibitem[{{Pols} {et~al.}(1995){Pols}, {Tout}, {Eggleton}, \&
  {Han}}]{chrisupdate}
{Pols}, O.~R., {Tout}, C.~A., {Eggleton}, P.~P., \& {Han}, Z. 1995, \mnras,
  274, 964, \dodoi{10.1093/mnras/274.3.964}

\bibitem[{{Pustylnik}(1998)}]{algol}
{Pustylnik}, I. 1998, Astronomical and Astrophysical Transactions, 15, 357,
  \dodoi{10.1080/10556799808201791}

\bibitem[{{Rasio} {et~al.}(1996){Rasio}, {Tout}, {Lubow}, \& {Livio}}]{rasio}
{Rasio}, F.~A., {Tout}, C.~A., {Lubow}, S.~H., \& {Livio}, M. 1996, \apj, 470,
  1187, \dodoi{10.1086/177941}

\bibitem[{{Rieutord} \& {Zahn}(1997)}]{tassoulbad}
{Rieutord}, M., \& {Zahn}, J.-P. 1997, ApJ, 474, 760, \dodoi{10.1086/303494}

\bibitem[{{Ruiter}(2020)}]{R2020}
{Ruiter}, A.~J. 2020, IAU Symposium, 357, 1, \dodoi{10.1017/S1743921320000587}

\bibitem[{{Ruiter} {et~al.}(2009){Ruiter}, {Belczynski}, \& {Fryer}}]{sne1a}
{Ruiter}, A.~J., {Belczynski}, K., \& {Fryer}, C. 2009, \apj, 699, 2026,
  \dodoi{10.1088/0004-637X/699/2/2026}

\bibitem[{{Schr{\"o}der} {et~al.}(1997){Schr{\"o}der}, {Pols}, \&
  {Eggleton}}]{starsdov}
{Schr{\"o}der}, K.-P., {Pols}, O.~R., \& {Eggleton}, P.~P. 1997, \mnras, 285,
  696, \dodoi{10.1093/mnras/285.4.696}

\bibitem[{{Schwarzschild} \& {H{\"a}rm}(1958)}]{schwarzscsemiconv}
{Schwarzschild}, M., \& {H{\"a}rm}, R. 1958, \apj, 128, 348,
  \dodoi{10.1086/146548}

\bibitem[{{Stancliffe} \& {Eldridge}(2009)}]{stancliffe09}
{Stancliffe}, R.~J., \& {Eldridge}, J.~J. 2009, \mnras, 396, 1699,
  \dodoi{10.1111/j.1365-2966.2009.14849.x}

\bibitem[{{Tassoul}(1987)}]{tassoul80s}
{Tassoul}, J.-L. 1987, ApJ, 322, 856, \dodoi{10.1086/165780}

\bibitem[{Terquem(2021)}]{terquem}
Terquem, C. 2021, Monthly Notices of the Royal Astronomical Society, 503,
  5789–5806, \dodoi{10.1093/mnras/stab224}

\bibitem[{{Terquem} \& {Martin}(2021)}]{termquemii}
{Terquem}, C., \& {Martin}, S. 2021, \mnras, 507, 4165,
  \dodoi{10.1093/mnras/stab2322}

\bibitem[{Vick \& Lai(2020)}]{vicklai}
Vick, M., \& Lai, D. 2020, Monthly Notices of the Royal Astronomical Society,
  496, 3767–3780, \dodoi{10.1093/mnras/staa1784}

\bibitem[{{von Zeipel}(1910)}]{vonzeipel}
{von Zeipel}, H. 1910, Astronomische Nachrichten, 183, 345,
  \dodoi{10.1002/asna.19091832202}

\bibitem[{{Wang} {et~al.}(2013){Wang}, {Justham}, \& {Han}}]{WJH2013}
{Wang}, B., {Justham}, S., \& {Han}, Z. 2013, A\&A, 559, A94,
  \dodoi{10.1051/0004-6361/201322298}

\bibitem[{{Webbink}(1984)}]{webbinksne}
{Webbink}, R.~F. 1984, \apj, 277, 355, \dodoi{10.1086/161701}

\bibitem[{{Woosley} \& {Kasen}(2011)}]{WK2011}
{Woosley}, S.~E., \& {Kasen}, D. 2011, \apj, 734, 38,
  \dodoi{10.1088/0004-637X/734/1/38}

\bibitem[{{Yoon} \& {Langer}(2004)}]{YL2004a}
{Yoon}, S.-C., \& {Langer}, N. 2004, \aap, 419, 623,
  \dodoi{10.1051/0004-6361:20035822}

\bibitem[{{Zahn}(1966)}]{zahn66}
{Zahn}, J.~P. 1966, Annales d'Astrophysique, 29, 313

\bibitem[{{Zahn}(1975)}]{zahn75}
{Zahn}, J.-P. 1975, A\&A, 41, 329

\bibitem[{{Zahn}(1977)}]{zahn77}
---. 1977, A\&A, 57, 383

\bibitem[{{Zahn}(1989)}]{zahnvisc}
{Zahn}, J.~P. 1989, \aap, 220, 112

\bibitem[{{Zimmerman} {et~al.}(2017){Zimmerman}, {Thompson}, {Mullally},
  {Fuller}, {Shporer}, \& {Hambleton}}]{pseudo}
{Zimmerman}, M.~K., {Thompson}, S.~E., {Mullally}, F., {et~al.} 2017, \apj,
  846, 147, \dodoi{10.3847/1538-4357/aa85e3}

\end{thebibliography}
\bibliographystyle{aasjournal}



\end{document}